\documentclass[seceq]{ptptex}

\newcommand{\EQ}{\begin{equation}}
\newcommand{\EN}{\end{equation}}
\newcommand{\bea}{\begin{eqnarray}}
\newcommand{\ena}{\end{eqnarray}}
\newcommand{\bdis}{\begin{displaymath}}
\newcommand{\edis}{\end{displaymath}}

\renewcommand{\a}{\alpha}
\renewcommand{\b}{\beta}

\renewcommand{\d}{\delta}

\renewcommand{\t}{\tau}

\newcommand{\dslash}{D\!\!\!\! \slash}
\newcommand{\caldslash}{{\cal D}\!\!\!\! \slash}
\newcommand{\pa}{\partial}

\newcommand{\nn}{\nonumber \\}

\title{
${\cal N}$=1 Supersymmetric Yang-Mills Theory \\
          in It${\bar {\rm o}}$ Calculus 
}

\author{
Naohito\ Nakazawa
}

\inst{
 High Energy Accelerator Research Organization (KEK) \\
     Tsukuba 305-0801, Japan
}

\abst{
The stochastic quantization method is applied to ${\cal N}$ = 1 supersymmetric Yang-Mills theory, in particular in 4 and 10 dimensions. In the 4 dimensional case, based on It${\bar {\rm o}}$ calculus, the Langevin equation is formulated in terms of the superfield formalism. The stochastic process manifestly preserves both the global ${\cal N}$ = 1 supersymmetry and the local gauge symmetry. The expectation values of the local gauge invariant observables in SYM$_4$ are reproduced in the equilibrium limit. In the superfield formalism, it is impossible in SQM to choose the so-called Wess-Zumino gauge in such a way to gauge away the auxiliary component fields in the vector multiplet, while it is shown that the time development of the auxiliary component fields is determined by the Langevin equations for the physical component fields of the vector multiplet in an \lq\lq almost Wess-Zumino gauge \rq\rq. The physical component expressions of the superfield Langevin equation are naturally extended to the 10 dimensional case, where the spinor field is Majorana-Weyl. By taking a naive zero volume limit of the ${\rm SYM}_{10}$, the IIB matirx model is studied in this context.
}

\begin{document}

\maketitle

\section{Introduction}

${\cal N}$ = 1 supersymmetric Yang-Mills theory in ten dimensions\cite{BSS} has attracted special interest in relation to superstring theories, not only because it describes a part of the low energy effective theory of superstrings\cite{GSO} but also because its reduction to one dimension describes an $N$ D-particle system as the light-cone eleven dimensional M-theory.\cite{IIA} In a naive zero volume limit with large $N$, it provides a constructive definition of type IIB superstring,\cite{IIB} where the supersymmetry is enhanced to ${\cal N}$ = 2, inherited from the Green-Schwarz superstring action. The constructive approach is supported by the fact that, due to the ${\cal N}$ = 2 supersymmetry, the light-cone IIB superstring field theory\cite{GSB} can be derived from the Schwinger-Dyson equation, or the so-called loop equation,\cite{MM} for Wilson loops in the IIB matrix model.\cite{IIB2} Since there is no manifestly Lorentz invariant formulation of IIB superstring field theory, the light-cone setting is necessary to investigate the equivalence. While if the large $N$ matrix model provides a constructive definition of type IIB superstring, we prefer to keep the manifest Lorentz covariance to obtain some hints for a manifestly Lorentz invariant description of superstring field theories. This motivates us to apply the stochastic quantization method (SQM)\cite{PW,DH} to large $N$ matrix models. As a significant advantage of SQM, it enables us to calculate the expectation values of gauge invariant observables, such as the Wilson loop, without the gauge fixing in terms of the Langevin equation that preserves the manifest Lorentz covariance. The application of SQM systematically incorporates a certain class of Schwinger-Dyson equations (loop equations) of the fundamental system. 
In the four-dimensional case,\cite{FZ} the non-perturbative effects in the supersymmetric Yang-Mills theory have been studied extensively in relation to the \lq\lq old fashioned \rq\rq matrix models in the context of the Dijkgraaf-Vafa theory,\cite{DV1,DV2} where the basic idea is also based on the analysis of the Schwinger-Dyson equation (loop equation). This also motivates us to apply SQM to the four-dimensional case in the superfield formalism. 

In this note, we consider stochastic quantization of ${\cal N}$ = 1 supersymmetric Yang-Mills theory (SYM$_d$) in $d$-dimensions, in particular $d = 4, 10$. Our main interest is to apply SQM to the IIB marix model. In a previous work,\cite{EKN}  we applied SQM to a prototype of reduced models, i.e., a naive zero volume limit of $U(N)$ Yang-Mills theory. We obtained a collective field theory of Wilson loops that resembles the covariantized bosonic part of the light-cone IIB superstring field theory. In order to develop the collective field theory approach in the context of SQM, we clarify the symmetry properties of the underlying stochastic process for SYM. SQM was first applied to supersymmetric models, free vector and chiral scalar fields, by Breit-Gupta-Zaks\cite{BGZ} in the superfield formalism. Ishikawa\cite{Ishikawa} extended its application to supersymmetric QED$_4$, incorporating the $U(1)$ local gauge invariance in this context. The regularization procedure for supersymmetric QED$_4$ was also discussed in Ref.\citen{Kalivas} in terms of the Schwinger-Dyson equation. In these works, it was clarified that the supersymmetric extension of the Langevin systems produces Langevin equations with kernels for auxiliary component fields and physical fermion fields.\cite{Sakita} In the application of SQM to SYM$_4$, it is necessary to incorporate both the non-abelian structure of the interaction and the global supersymmetry. For SYM$_{10}$, the basic problem is to apply SQM to the Majorana-Weyl fermion. In order to make manifest the covariance of the Langevin equations both under the local gauge transformation and under the global supersymmetry transformation, we first apply SQM to $SU(N)$ SYM$_{4}$ in the superfield formalism.\cite{FWZ-SS} Using the superfield Langevin equation, the expectation values of local gauge invariant observables for SYM$_4$ are reproduced in the equilibrium limit. We also derive the corresponding Fokker-Planck equation written in terms of the superfield, which implies that the probability distribution reproduces that of SYM$_4$ formally in the equilibrium limit. It is shown that the superfield Langevin equation manifestly preserves both the local gauge symmetry and the global ${\cal N}$ = 1 supersymmetry in the sense of It${\bar {\rm o}}$. Its geometrical meaning is also clarified in superspace by introducing a nontrivial path-integral measure that defines a local gauge invariant partition function in the superfield formalism. Next, we extend it to the ten-dimensional case. Since there is no superfield formalism for SYM$_{10}$, the Langevin equations for $SU(N)$ SYM$_4$ expressed in terms of physical component fields $( \lambda, {\bar \lambda}, v_m )$ are naturally extended to those for $SU(N)$ SYM$_{10}$ by introducing a chiral projection for the two Majorana noise variables. Then, we study the IIB matrix model by taking the naive zero volume limit of $U(N)$ SYM$_{10}$. The Fokker-Planck Hamiltonian is also derived for the purpose of constructing a collective field theory of Wilson loops in this context. 

The paper is organized as follows. In \S 2, we formulate the Langevin equation and the Fokker-Planck equation for SYM$_4$ in the superfield formalism. In \S 3, we decompose the superfield Langevin equation into its components. We introduce an \lq\lq almost Wess-Zumino gauge\rq\rq to gauge away the auxiliary fields in the vector superfield. In a precise sense, the Wess-Zumino gauge fixing is impossible for the superfield Langevin equation. However, we show that the time development of the auxiliary fields is completely determined by only the component fields in the Wess-Zumino gauge. Section 4 is devoted to the Euclidean formulation for component Langevin equations. We extend the formulation in component fields for SYM$_4$ to SYM$_{10}$ in \S 5. The IIB matrix model is discussed in \S 6. In Appendix A, we introduce a local gauge invariant definition of the partition function in the superfield formalism to derive the Fokker-Planck equation. In particular, we demonstrate the local gauge invariance of the superspace path-integral measure. Appendix B is devoted to a derivation of the supersymmetric kernel for the superfield Langevin equation. The gauge fixing problem is also discussed there and we derive the superfield Langevin equation in the \lq\lq almost Wess-Zumino gauge\rq\rq. In Appendix C, we derive the component expressions for the superfield Langevin equation. 


\section{SYM$_4$ in SQM}

We apply SQM to (SYM)$_4$ in the superfield formalism to make its covariance manifest. Although we work in the Minkowski space in the following two sections, it is always possible to perform the Wick rotation to the Euclidean space for the convergence of the Langevin equation, then back to the Minkowski space by analytic continuation.\cite{BGZ} The Euclidean prescription is discussed in \S 4. 

A vector superfield $V = V^\dagger$ is given by 
\bea
\label{eq:superfield}
V (x,\ \theta,\ {\bar \theta})
& = & 
C + i\theta\chi - i{\bar \theta}{\bar \chi} 
+ \displaystyle{\frac{i}{2}}\theta^2 [ M + i N ] 
- \displaystyle{\frac{i}{2}}{\bar \theta}^2 [ M - i N ] 
- \theta\sigma^m {\bar \theta} v_m  \nn
& + & 
i \theta^2 {\bar \theta} [ {\bar \lambda} 
        + \displaystyle{\frac{i}{2}} {\bar \sigma}^m \pa_m \chi ] 
- i {\bar \theta}^2 \theta [ \lambda 
        + \displaystyle{\frac{i}{2}} \sigma^m \pa_m {\bar \chi} ] 
+ \displaystyle{\frac{1}{2}}\theta^2 {\bar \theta}^2 [ D
        + \displaystyle{\frac{1}{2}} \pa^2 C]     ,
\ena
where the superfield $V$ and its component fields are $SU(N)$ algebra-valued: $V = V^a t_a$ with $t_a \in su(N)$, which satisfy $[ t_a, t_b] = if^{abc} t_c$ and ${\rm Tr}( t_a t_b ) = {\rm k}\delta_{ab}$.  
We use a notation closely related to that in Ref.~\citen{WB}. The transformation property of the vector superfield under the local gauge transformation is given by 
\bea
\label{eq:local-gauge-tr.}
e^{2V} 
 \rightarrow  
e^{-i\Lambda^\dagger} e^{2V}e^{i\Lambda}     \ , 
\ena
where $\Lambda$ and $\Lambda^\dagger$ are $SU(N)$ algebra-valued chiral superfields which satisfy 
$
{\bar D}_{\dot \alpha} \Lambda = D_\alpha \Lambda^\dagger = 0 \ . 
$
By introducing a local gauge covariant chiral superfield $W_\alpha$, the so-called \lq\lq Gluino field\rq\rq, and its conjugation ${\bar W}_{\dot \alpha}$, 
\bea
\label{eq:covariant-fiel-strength}
W_\alpha 
& = & 
- \displaystyle{\frac{1}{8}} {\bar D}^2 e^{-2V}D_\alpha e^{2V} 
     \ , \nn 
{\bar W}_{\dot \alpha} 
& = & 
 \displaystyle{\frac{1}{8}} D^2 e^{2V}{\bar D}_{\dot \alpha} e^{-2V}            \ , 
\ena
the action of SYM$_4$ is given by 
\bea
\label{eq:4Daction}
S 
& = & 
\int d^4x d^2\theta d^2{\bar \theta} \displaystyle{\frac{1}{4{\rm k}g^2}} {\rm Tr}    \Big(
W^\alpha W_\alpha \delta^2 ( {\bar \theta} ) + {\bar W}_{\dot \alpha}{\bar W}^{\dot \alpha} \delta^2 ( \theta )
\Big)           \  , \nn 
& = & 
\displaystyle{\frac{1}{{\rm k}g^2}}  \int d^4x {\rm Tr} \Big( 
- \displaystyle{\frac{1}{4}} v_{mn} v^{mn} 
- \displaystyle{\frac{i}{2}} \lambda \sigma^m {\cal D}_m {\bar \lambda} 
- \displaystyle{\frac{i}{2}} {\bar \lambda} {\bar \sigma}^m {\cal D}_m \lambda 
+ \displaystyle{\frac{1}{2}} D^2         
\Big) \ . 
\ena
Here, the expression with component fields is written in the Wess-Zumino gauge,  
$
v_{mn} = \pa_m v_n - \pa_n v_m + i \Big[ v_m,\ v_n \Big]  
$
 and 
$
{\cal D}_m {\lambda} = \pa_m {\lambda} + i \Big[ v_m,\ {\lambda} \Big]  . \ 
$
In the following, we do $not$ assume the Wess-Zumino gauge.

We define the time evolution of the vector superfield, 
$V( \t+\Delta \t ) \equiv V ( \t ) + \Delta V( \t )$, with respect to the stochastic time $\t$ in terms of It${\bar {\rm o}}$ calculus.\cite{Ito} We use discretized notation to allow a clear understanding. The Langevin equation is defined from the Fokker-Planck equation that reproduces the probability distribution $e^{-S}$ with $S$ in (\ref{eq:4Daction}) in the equilibrium limit for 
$\t \rightarrow \infty$. 
First, we regard $e^{2V}$ as a fundamental variable, because of its simple transformation property. Then we derive the Langevin equation for the vector superfield $V$. 

To derive the superfield Langevin equation, we assume the following form for the time evolution of $e^{2V}$:   
$
e^{2V}( \t + \Delta \t ) 
=
 e^{2V}( \t ) + \Delta e^{2V}( \t )
$. 
\bea
\label{eq:4DLangevin-eq1}
e^{2V} ( \t + \Delta \t, z ) 
& = & 
e^{- \Delta \t X ( \t, z ) +  \Delta w ( \t, z )} e^{2V} ( \t, z )    \ , \nn 
\langle 
\Delta w_{ij} ( \t, z ) \Delta w_{kl} ( \t, z' ) 
\rangle_{\Delta w_\t}   
& = &
2{\rm k} \Delta \t \Big( 
\delta_{il}\delta_{jk} - \displaystyle{\frac{1}{N}}\delta_{ij}\delta_{kl} 
\Big) \delta ( z - z' )        \ , \nn 
\delta ( z - z' ) 
& \equiv &
\delta^2 ( \theta - \theta' ) 
\delta^2 ( {\bar \theta} - {\bar \theta}' ) \delta^4 ( x - x' )   \ , 
\ena
where $z$ denotes the set of coordinates $( x, \theta, {\bar \theta} )$ in superspace, $\Delta w ( \t, z )$ is a noise superfield, and  
$X ( \t, z )$ is determined from the Fokker-Planck equation so as to reproduce the superfield SYM$_4$.
We assume that $X$ and $\Delta w$ are $SU(N)$ algebra-valued: \ 
$X = X_a t^a$ and 
$\Delta w = \Delta w_a t^a$. 
In the following, $\langle ... \rangle_{\Delta w_\t}$  indicates that the expectation value is evaluated by means of the noise correlation at the stochastic time $\t$ defined in (\ref{eq:4DLangevin-eq1}).
Up to order $\Delta \t$, we obtain
\bea
\label{eq:4DLangevin-eq2}
\Delta e^{2V} ( \t, z ) 
=  
\Big( 
- \Delta \t X ( \t, z ) +  \Delta w ( \t, z ) 
\Big) e^{2V} ( \t, z )    
       \ . 
\ena

We now comment on the contact term which would appear on the r.h.s. of the Langevin equation (\ref{eq:4DLangevin-eq2}) in It${\bar {\rm o}}$ calculus. It can be evaluated from the noise correlation in (\ref{eq:4DLangevin-eq1}) as 
\bea
\label{eq:4DLangevin-contact-term}
& {} & \langle
\displaystyle{\frac{1}{2}} \Delta w ( \t, z )_{ik} \Delta w ( \t, z )_{kj} e^{2V} ( \t, z )   
\rangle_{\Delta w_\t}    \nn
& {} & \qquad\qquad\qquad\qquad 
 = 
{\rm k}\Delta \t \displaystyle{\frac{N^2 - 1}{N}} \delta_{ij} 
\lim_{z' \rightarrow z} \delta ( z - z' ) e^{2V} ( \t, z )   \  .
\ena
To evaluate this term precisely, we might need a regularization procedure. The regularization for the $\delta$-function, $\delta^4 ( x - x' )$, can be realized by smearing it, while the relation  
$
\lim_{\theta' \rightarrow \theta,\ {\bar \theta}' \rightarrow {\bar \theta}} \delta^2 ( \theta - \theta' ) \delta^2 ( {\bar \theta} - {\bar \theta}' ) = 0
$ 
 is a direct consequence of the supersymmetry. We thus, as a working hypothesis to derive the basic Langevin equation for SYM$_4$, ignore the contribution which comes from this type of ill-defined contact terms.\footnote{As a regularization procedure, the prescription 
$
\delta^4(0) \delta^2 ( \theta - \theta' ) \delta^2 ( {\bar \theta} - {\bar \theta}' )|_{\theta'=\theta,\ {\bar \theta}'={\bar \theta}} = 0 
$ 
is consistent with regularization by dimensional reduction.\cite{Siegel-CJN}} In non-supersymmetric theories, such as lattice gauge theories, this type of contact terms, in general, plays a particular role for the general coordinate invariance of the Langevin equation in It${\bar {\rm o}}$ calculus.\cite{Nakazawa1} This type of contact terms is also the origin of anomalies in the context of SQM. We discuss this point later.
 
For the time evolution, we require that 
\bea
\label{eq:constraint-eq1}
V ( \t + \Delta \t)^\dagger = V ( \t + \Delta \t )  \ , \quad
{\rm Tr} V ( \t + \Delta \t) = 0 \ .   
\ena
The hermiticity of the vector superfield is ensured as follows. 
Let us rewrite the Langevin equation (\ref{eq:4DLangevin-eq1}) as 
\bea
\label{eq:constraint-eq2}
& {} & e^{2V} ( \t + \Delta \t )    \nn 
& = & 
e^{ {1\over2}( - \Delta \t X ( \t ) +  \Delta w ( \t )  ) } e^{2V} ( \t ) e^{ {1\over2}( 
- \Delta \t e^{-2V} X ( \t ) e^{2V} 
+ e^{-2V} \Delta w ( \t ) e^{2V} )  }     .   
\ena
This expression implies that the $SU(N)$ algebra-valued quantities $X$ and $\Delta w$ must satisfy the hermiticity conditions 
\bea
\label{eq:constraint-eq3} 
X ( \t )^\dagger 
& = & e^{-2V} X ( \t ) e^{2V}    \ , \nn
\Delta w( \t )^\dagger 
& = &  e^{-2V} \Delta w ( \t ) e^{2V}   \ .
\ena
We choose $X$ so that the first constraint is satisfied. The hermiticity assignment (\ref{eq:constraint-eq3}) for the noise superfield defines the hermitian conjugation of its components, which implies that the noise superfield, $\Delta w$, is not a vector superfield. 
From the definition of the hermitian conjugation for the noise superfield, we obtain their correlations, 
\bea
\label{eq:noise-correlation-eq1}
\langle 
\Delta w_{ij}^\dagger ( \t, z ) \Delta w_{kl}^\dagger ( \t, z' ) 
\rangle_{\Delta w_\t}   
& = &
2{\rm k} \Delta \t \Big( 
\delta_{il}\delta_{jk} - \displaystyle{\frac{1}{N}}\delta_{ij}\delta_{kl} 
\Big) \delta ( z - z' )        \ , \nn 
\langle 
\Delta w_{ij} ( \t, z ) \Delta w_{kl}^\dagger ( \t, z' ) 
\rangle_{\Delta w_\t}   
& = &
2{\rm k} \Delta \t \Big( 
(e^{2V})_{il} (e^{-2V})_{kj} - \displaystyle{\frac{1}{N}} \delta_{ij} \delta_{kl} 
\Big) \delta ( z - z' )        \ .
\ena
We note that the noise correlation (\ref{eq:noise-correlation-eq1}) is not the definition of the hermitian conjugate $\Delta w^\dagger$, but it follows from (\ref{eq:4DLangevin-eq1}) and (\ref{eq:constraint-eq3}). The definition of the hermitian conjugation of the noise superfield (\ref{eq:constraint-eq3}) is realized by determining the conjugate component fields so as to satisfy (\ref{eq:constraint-eq3}), which is possible because the noise superfield is not a vector superfield. Therefore, it is realized only by the white (Gaussian) noise correlation defined in (\ref{eq:4DLangevin-eq1}). In particular, in the sense of It${\bar {\rm o}}$ calculus, the noise variable is not correlated with the equal stochastic time dynamical variables. This means that the r.h.s of the noise correlation in (\ref{eq:noise-correlation-eq1}) is not the expectation value. 

From (\ref{eq:constraint-eq2}), we obtain a form of the Langevin equation in which the hermiticity of the vector superfield is manifest throughout the time development: 
\bea
\label{eq:4DLangevin-eq3} 
\Delta e^{2V} 
& = & 
\displaystyle{\frac{1}{2}}\Big( - \Delta \t X ( \t ) +  \Delta w ( \t )  \Big) e^{2V} ( \t )    \nn
& {} & \qquad +  e^{2V} ( \t ) \displaystyle{\frac{1}{2}}\Big( - \Delta \t X ( \t )^\dagger +  \Delta w ( \t )^\dagger  \Big)      \ . 
\ena
Since this Langevin equation is equivalent to (\ref{eq:4DLangevin-eq2}), we consider (\ref{eq:4DLangevin-eq2}) in the following for simplicity. 

Before we derive the corresponding Fokker-Planck equation and determine the explicit form of the quantity $X$, we comment on the transformation property of the Langevin equation (\ref{eq:4DLangevin-eq2}) under the stochastic-time $independent$ local gauge transformation, which means that the transformation parameters, $\Lambda$ and $\Lambda^\dagger$ in (\ref{eq:local-gauge-tr.}), do not depend on the stochastic time. The l.h.s. of (\ref{eq:4DLangevin-eq2}) is transformed as 
$
 \Delta e^{2V} 
\rightarrow e^{-i\Lambda\dagger} ( \Delta e^{2V} ) e^{i\Lambda}   \ .
$
For the covariance of the Langevin equation, we require 
\bea
\label{eq:noise-transformation}
X 
& \rightarrow & 
e^{-i\Lambda^\dagger} X e^{i\Lambda^\dagger} \ , \nn
\Delta w
& \rightarrow & 
e^{-i\Lambda^\dagger} \Delta w e^{i\Lambda^\dagger} \ . 
\ena
We note that the noise correlation in (\ref{eq:4DLangevin-eq1}) and the first equation in (\ref{eq:noise-correlation-eq1}) are $invariant$ while the second equation in (\ref{eq:noise-correlation-eq1}) is $covariant$ under the transformation (\ref{eq:noise-transformation}). The hermiticity assignment in (\ref{eq:constraint-eq3}) is also consistent with the transformation property. Hence, we conclude that the Langevin equation and the noise correlation manifestly preserve the global ${\cal N}$ = 1 supersymmetry, as well as the local gauge symmetry, if we obtain $X$ in the Langevin equation which satisfies the hermiticity condition and the transformation property in (\ref{eq:constraint-eq3}) and (\ref{eq:noise-transformation}), respectively. 

The Fokker-Planck equation which corresponds to the Langevin equation (\ref{eq:4DLangevin-eq2}) is given by (see Appendix A) 
\bea
\label{eq:Fokker-Planck-eq1}
\displaystyle{\frac{\pa\ }{\pa \t}}P( \t, e^{2V} ) 
= 
\int\!\!\! dz {\hat {\cal E}}_a(z) \Big(  
{\hat {\cal E}}^a (z) + X (z)^a  
\Big) P( \t, e^{2V} )         \ . 
\ena 
Here ${\hat {\cal E}}_a (z)$ has been introduced as an analogue of the left Lie derivative on a group manifold: 
\bea
\label{eq:left-derivative-analogue} 
{\hat {\cal E}}_a (z) 
& \equiv & 
\displaystyle{\frac{1}{2}}
 L_a^{\ b} (z) \displaystyle{\frac{\delta\ \ }{\delta V^b (z)}}   \ , \nn 
K_a^{\ b} (z)
& \equiv & 
\displaystyle{\frac{1}{{\rm k}}}\int^1_0 ds {\rm Tr}\Big( 
e^{2sV (z)}t_a e^{-2sV (z)} 
\Big)                                                             \ , \nn 
K_a^{\ c} (z) L_c^{\ b} (z) 
& = & 
L_a^{\ c} (z) K_c^{\ b} (z) = \delta_a^{\ b}      \ . 
\ena 
It satisfies the relation 
$
{\hat {\cal E}}_a (z) e^{2V (z')} = t_a e^{2V (z')}\delta ( z -z' )
$. 
The local gauge invariant definition of the expectation value of an observable 
$F [ e^{2V} ]$ is given by 
\bea
\label{eq:super-gauge-inv.-expectation}
\langle 
F [ e^{2V} ] ( \t ) 
\rangle_{\Delta w} 
\equiv \int\!\!\! F [ e^{2V} ] P( \t , e^{2V} ) \sqrt{G} {\cal D}V  \ .
\ena
Here $\langle {...} \rangle_{\Delta w}$ indicates that the expectation value is evaluated with respect to all the noise for 
$\t' \le \t - \Delta \t$. 
On the r.h.s., we have introduced the metric, 
$ G^{ab} \equiv {1\over 4}L_c^{\ a} L_c^{\ b}$ and 
$ G_{ab} \equiv 4K_a^{\ c} K_b^{\ c}$,  
to define a local gauge invariant superspace path-integral measure, 
$\sqrt{G}{\cal D}V$. 
Here, 
$G$ denotes ${\rm det}G_{ab}$. (See Appendix A for details.)  
To reproduce the superfield action for ${\cal N}$=1 SYM$_4$ in the equilibrium limit, we simply choose 
\bea
\label{eq:Fokker-Planck-eq2}
X (z) 
& \equiv & 
{\hat {\cal E}} (z) S         \ , \nn 
& = & 
{\rm k} \Big( 
e^{2V} \displaystyle{\frac{\delta S}{\delta (e^{2V })^t}}  
- \displaystyle{\frac{1}{N}} {\bf 1 } {\rm Tr} 
( e^{2V} \displaystyle{\frac{\delta S}{\delta ( e^{2V } )^t}} ) 
\Big)                  \ , \nn
& = & 
- \displaystyle{\frac{1}{4g^2}}\Big( 
 e^{2V} 
{\cal D}^\alpha W_\alpha e^{-2V}   
+  
{\bar {\cal D}}_{\dot \alpha} {\bar W}^{\dot \alpha} 
\Big)   \ , 
\ena
where
\bea
{\cal D}^\alpha W_\alpha 
& \equiv & 
D^\alpha W_\alpha + \Big\{ 
W_\alpha , e^{-2V} D^\alpha e^{2V}
\Big\}                 \ , \nn
{\bar {\cal D}}_{\dot \alpha} {\bar W}^{\dot \alpha} 
& \equiv & 
{\bar D}_{\dot \alpha} {\bar W}^{\dot \alpha} + \Big\{ 
{\bar W}^{\dot \alpha} , e^{2V} {\bar D}_{\dot \alpha} e^{-2V} 
\Big\}   \ .
\ena
This satisfies the hermiticity condition (\ref{eq:constraint-eq3}) and the transformation property (\ref{eq:noise-transformation}). We note that, in the Wess-Zumino gauge, the two terms $W^\alpha W_\alpha$ and ${\bar W}_{\dot \alpha}{\bar W}^{\dot \alpha}$ in the superfield Lagrangian (\ref{eq:4Daction}) are equivalent, up to an imaginary total divergence term which does not contribute to the equations of motion. In fact, the reality condition 
$
\int \!\! d^4 x d^2\theta {\rm Tr} W^{\a}W_{\a} 
= \int \!\! d^4 x d^2{\bar \theta} {\rm Tr} {\bar W}_{\dot \a}{\bar W}^{\dot a} 
$ 
implies that 
$
e^{2V} {\cal D}^\a W_\a e^{-2V} 
= {\bar {\cal D}}_{\dot a} {\bar W}^{\dot a} \ .
$
For the manifest hermiticity, we employ both terms. For the perturbative analysis, we can discard one of them in the basic Langevin equation. 

We thus obtain the Langevin equation for $e^{2V}$, 
\bea
\label{eq:4DLangevin-eq4}
( \Delta e^{2V} )e^{-2V} ( \t, z )
=  
- \Delta \t {\hat {\cal E}} ( \t, z ) S + \Delta w ( \t, z )       \ , 
\ena 
where 
${\hat {\cal E}} ( \t, z ) = t_a {\hat {\cal E}}^a ( \t, z ) $. 
We also comment that the Fokker-Planck equation, 
\bea
\label{eq:4DFokker-Planck-eq2}
\displaystyle{\frac{\pa}{\pa \t}} P( \t ) 
= 
\displaystyle{\frac{1}{{\rm k}}}\int\!\!\! dz {\rm Tr} {\hat {\cal E}} (z) 
\Big( 
{\hat {\cal E}} (z) + ( {\hat {\cal E}}(z) S )
\Big) P( \t )           \ , 
\ena
is invariant under the local gauge transformation. In the equilibrium limit, this indicates that the probability distribution behaves formally as 
$P( \t , e^{2V} ) \rightarrow e^{- S}$. 
In a precise sense, what we have shown is that $e^{-S}$ is a stationary solution of the Fokker-Planck equation (\ref{eq:4DFokker-Planck-eq2}). This ensures that $the$ $expectation$ $values$ $of$ $local$ $gauge$ 
$invariant$ $observables$ for SYM$_4$ are reproduced in the equilibrium limit. This does not imply, however, the existence of a unique equilibrium distribution, which can be proven only for a smooth positive semi-definite stationary solution describing a finite number of degrees of freedom. In this respect, to prove the existence of a unique equilibrium distribution for SYM$_4$, we need a regularization procedure to reduce the number of degrees of freedom that preserves the global supersymmetry as well as the local gauge symmetry. It is also necessary to introduce a so-called stochastic gauge fixing procedure to give a drift force to the longitudinal degrees of freedom. We discuss the gauge fixing procedure in the next section. 
Equation (\ref{eq:super-gauge-inv.-expectation}) also clearly shows that we need the path-integral measure $\sqrt{G} {\cal D}V$ even in the equilibrium limit for the definition of the local gauge invariant partition function in the path-integral formalism.

Now we derive the Langevin equation for the vector superfield $V (z)$. To do this, we evaluate the r.h.s. of (\ref{eq:4DLangevin-eq4}). We find  
\bea
\label{eq:contact-term-eq1}
( \Delta U ) U^{-1} 
 =  
\Delta V^a (\pa_a U ) U^{-1} + {1\over 2}\Delta V^a \Delta V^b ( \pa_a \pa_b U ) U^{-1}  + O(\Delta\t^{3/2})          \ , 
\ena
where $U \equiv e^{2V}$. We again ignore the contribution from the contact term. Then, (\ref{eq:contact-term-eq1}) reads 
\bea
\label{eq:4DLangevin-eq6}
\Delta V^a ( \t, z)
 = 
 \displaystyle{\frac{1}{2}} 
 \Big( - \Delta \t {\hat {\cal E}}( \t, z ) S + \Delta w ( \t, z )  \Big) ^b L_b^{\ a} ( \t, z )          \ .
\ena

The construction of the superfield Langevin equation (\ref{eq:4DLangevin-eq4}) and the Fokker-Planck equation (\ref{eq:4DFokker-Planck-eq2}) is a direct analogue of that in the lattice gauge theories, where the underlying stochastic process is well established.\cite{DDH-GL-H, Sakita} 
 As mentioned above, 
the contact term contribution, which we have ignored here not because it is ill-defined but because of the supersymmetry requirement, plays an essential role for the covariant nature of the Langevin equation in lattice gauge theories.\cite{Nakazawa1} In a lattice gauge theory, because the link variable is a group element, the stochastic process and the integration are defined on the group manifold. The geometry of the group manifold in the lattice gauge theory is reflected in general coordinate invariant forms of the Langevin equation and the Fokker-Planck equation in the sense of It${\bar {\rm o}}$ calculus. However, in case of SYM$_4$ in the superfield formalism, the fundamental variable $U = e^{2V}$ is not a group element, and the requirement is too strong to give a well-defined Langevin equation. As we explain in Appendix A, we need only the invariance of the partition function under the local gauge transformation of the superfield. Since the local gauge transformation is interpreted as a Killing vector on the superspace, it is not necessary to require the general coordinate invariant form under arbitrary general coordinate transformations for the covariance of the Langevin equation and the corresponding Fokker-Planck equation.     

The geometrical interpretation of the covariant nature of the superfield Langevin equation (\ref{eq:4DLangevin-eq6}) is the following. In the derivation of (\ref{eq:4DLangevin-eq6}), one would define a covariant form of $\Delta V^a$, 
$
\label{eq:covariant-derivative}
\Delta_{\rm cov} V^a \equiv \Delta V^a 
+ \Delta \t \Gamma^a_{\ bc}G^{bc} 
$ 
.\cite{Graham}  The local gauge transformation of the second term of this expression, 
$
\Delta \t \Gamma^a_{\ bc}G^{bc}  
=
- \Delta \t 
G^{-{1\over2}} \pa_b ( G^{1\over2} G^{ab} )    \ , 
$
would yield a contact term and it 
 would cancel the non-covariant contact term that would appear in the transformation of the time development of $\Delta V ( \t, z )$: 
\bea
\label{eq:local-gauge-tr-eq2}
\delta ( \Delta V^a (z) )
& = & 
\int \!\!\! dz' \Delta V^b (z') \displaystyle{\frac{\pa \delta V^a (z)}{\pa V^b (z')}}     \nn
& {} & \qquad + 
\int \!\!\! dz' dz'' \displaystyle{\frac{1}{2}}\Delta V^b (z') \Delta V^c (z'') \displaystyle{\frac{\pa^2 \delta V^a (z)}{\pa V^b (z') \pa V^c (z'')}}    \ , \nn 
& = &
-  \displaystyle{\frac{i}{2}}( \Lambda^\dagger (z))^b \pa_c L_b^{\ a} (z) \Delta V^c (z)
+  \displaystyle{\frac{i}{2}} \pa _c L^a_{\ b} (z) \Lambda^b (z) \Delta V^c (z)          \nn 
 & {} & \qquad + 
( {\rm contact\ term} )        \ .
\ena
Here, since we have ignored the contact term contribution, it is not necessary to introduce a general coordinate invariant expression. 
 The covariance is manifest if we simply write the Langevin equation (\ref{eq:4DLangevin-eq6}) and the noise correlation as 
\bea
\label{eq:4DLangevin-eq7}
\Delta V^a ( \t, z )
& = & 
-\Delta \t G^{ab} ( \t, z ) \displaystyle{\frac{\delta S}{\delta V^b (z)}} 
+ \Delta_w \Xi^a ( \t, z )    \ , \nn
\langle 
\Delta_w \Xi^a ( \t, z ) \Delta_w \Xi^b ( \t, z' ) 
\rangle_{\Delta w_\t} 
& = & 
2 \Delta \t G^{ab} ( \t, z )  \delta ( z - z' )   , 
\ena
where we have introduced a collective noise superfield, 
$ \Delta_w \Xi^a ( \t, z ) = {1\over 2}\Delta w^b ( \t, z ) L_b^{\ a} ( \t, z ) $. $G^{ab} ( \t, z )$ can be regarded as a kernel written in the superfield (see Appendix B). This is one of the main results of this paper. We note that, in the collective noise correlation in (\ref{eq:4DLangevin-eq7}), the r.h.s. is not the expectation value, because in It${\bar {\rm o}}$ calculus, the noise variable is uncorrelated with the equal stochastic-time dynamical variables. This property is essential to construct a collective field theory in the SQM approach.  
The corresponding Fokker-Planck equation written in terms of the superfield is given by 
\bea 
\label{eq:4DFokker-Planck-eq3}
\displaystyle{\frac{\pa}{\pa \t}} P( \t, V ) 
= 
\int\!\!\! dz \displaystyle{\frac{\delta\ }{\delta V^a (z)}}  \Big\{ 
G^{ab}( \t, z ) \Big( 
\displaystyle{\frac{\delta\ }{\delta V^b (z)}} + 
 \displaystyle{\frac{\delta S}{\delta V^b (z)}} 
 \Big)
 P( \t, V )  \Big\}         \ ,
\ena
where $P( \t, V )$ is a scalar probability, defined in (\ref{eq:super-gauge-inv.-expectation}). 
As an application, with the Langevin equation (\ref{eq:4DLangevin-eq7}), we can derive the supersymmetric Schwinger-Dyson equation for Wilson loops (or the supersymmetric loop equation) written in the superfield formalism, which has already been found in Ref.~\citen{IT}. 

Although we have postponed the explicit evaluation of the contact term (\ref{eq:4DLangevin-contact-term}), it could be done by using the background field method with a regularization procedure that preserves the global supersymmetry and the background local gauge symmetry. If the evaluation successfully extracted a non-trivial contribution from the contact term, as has been precisely shown in the case of lattice gauge theories\cite{Nakazawa1}, we would obtain, in a formal sense, the Langevin equation and the Fokker-Planck equation 
\bea
\label{eq:Would-be-4DLangevin-eq}
\Delta V^a ( \t, z )
& = & 
-\Delta \t G^{ab} ( \t, z ) \displaystyle{\frac{\delta S}{\delta V^b (z)}} 
- \Delta \t \Gamma^a_{\ bc}G^{bc} ( \t, z )
+ \Delta_w \Xi^a ( \t, z )    \ , \nn 
\displaystyle{\frac{\pa}{\pa \t}} P( \t ) 
& = & 
\int\!\!\! dz \displaystyle{\frac{1}{\sqrt{G}}}
\displaystyle{\frac{\delta\ }{\delta V^a (z)}} 
 \Big\{ \sqrt{G}
G^{ab}( \t, z ) \Big( 
\displaystyle{\frac{\delta\ }{\delta V^b (z)}} + 
 \displaystyle{\frac{\delta S}{\delta V^b (z)}} 
 \Big)
 P( \t )  \Big\}         \ .
\ena
In contrast to the lattice gauge theory case, the second term on the r.h.s. of the Langevin equation is ill-defined without the regularization procedure, due to  a $\delta (z-z')|_{z'=z}$ type  singularity. One way to solve this problem is to introduce an explicit regularization procedure applied to the noise correlation defined in (\ref{eq:4DLangevin-eq1}). Although a prototype of such a regularization procedure has been discussed,\cite{BGZ} we do not consider such a possibility in this note.


\section{Superfield Langevin equation in component fields}

In this section, we consider the component expansion of the superfield Langevin equation (\ref{eq:4DLangevin-eq6}) [or equivalently, (\ref{eq:4DLangevin-eq7})]. In the application of SQM to the free supersymmetric vector model (the abelian limit of SYM$_4$) in the superfield formalism, Breit, Gupta and Zaks\cite{BGZ} showed that it is necessary to perform the component expansion without choosing a special gauge, such as the Wess-Zumino gauge. In general, the Langevin equations for physical components, ($v_m$, $\lambda$, ${\bar \lambda}$, $D$), include the stochastic time evolution for the auxiliary components, ($C$, $\chi$, ${\bar \chi}$, $M$, $N$). In order to eliminate these auxiliary fields, one needs the Langevin equations for the auxiliary fields, ($C$, $\chi$, ${\bar \chi}$, $M$, $N$).  This means that the auxiliary fields cannot be gauged away in the superfield Langevin equation by requiring the Wess-Zumino gauge. Instead, they are eliminated by using their own Langevin equations. In view of the transformation properties of the superfield Langevin equation, this situation results from the fact that the variation with respect to the stochastic time is $not\ covariant$ under the stochastic-time $dependent$ local gauge transformation. As we showed at the end of the previous section, the superfield Langevin equation is covariant under the stochastic-time $independent$ local gauge transformation for which the transformation parameter, i.e. a chiral superfield $\Lambda ( z )$, is stochastic-time independent. While we need a stochastic-time $dependent$ local gauge transformation for which the transformation parameter $\Lambda ( \t, z )$ is stochastic-time $dependent$ in order to choose the Wess-Zumino gauge in the superfield Langevin equation. Therefore, we need a component expansion without the Wess-Zumino gauge fixing. A non gauge fixing analysis is also necessary to show explicitly the role of the auxiliary fields, in particular $C$, in the superspace path-integral measure defined in Appendix A. Unfortunately, an explicit component expansion without the Wess-Zumino gauge is extremely tedious. Instead of doing this, we derive the superfield Langevin equation in an \lq\lq almost Wess-Zumino gauge\rq\rq, in which the auxiliary fields appear only through their derivatives with respect to the stochastic time. This is achieved through a redefinition of the vector superfield with the local gauge transformation (\ref{eq:Wess-Zumino-eq1}) that does not change the component degrees of freedom. Under the transformation (\ref{eq:Wess-Zumino-eq1}), the action of SYM$_4$ is transformed to the expression (\ref{eq:4Daction}) in the Wess-Zumino gauge.  While in SQM, all the auxiliary component fields remain and it is impossible to gauge away the auxiliary component fields.  Since the auxiliary fields appear in the superfield Langevin equation only through their derivatives with respect to the stochastic time, the time development of the auxiliary fields is completely determined by the values of the physical component fields, ($v_m$, $\lambda$, ${\bar \lambda}$, $D$), and the noise component fields in the \lq\lq almost Wess-Zumino gauge\rq\rq. A detailed derivation demonstrating this fact is given in Appendix B.

By introducing the Lie derivative $L_V$ defined by $ L_V X \equiv [V, X]$,\cite{GGRS} the superfield Langevin equation (\ref{eq:4DLangevin-eq6}) (or equivalently (\ref{eq:4DLangevin-eq7})) can be expressed as 
\bea
\label{eq:4DLangevin-eq8}
\Delta V ( \t, z )
& = & 
 \Delta \t \displaystyle{\frac{1}{4g^2}}  \Big( 
( 1 - e^{-2L_V} )^{-1} L_V {\cal D}{\cal W}      \nn
& {} & \qquad \qquad \qquad 
+ (e^{2L_V} - 1 )^{-1} L_V {\bar {\cal D}}{\bar {\cal W}} 
\Big) + \Delta_w \Xi ( \t, z ) \ ,  \nn 
\Delta_w \Xi ( \t, z ) 
& \equiv & 
( e^{2L_V} - 1 )^{-1} L_V \Delta w  ( \t, z )    \ .
\ena

We introduce a superfield $V |_{{\rm WZ}} ( \t, z )$, which consists of only the component fields in the Wess-Zumino gauge. The auxiliary degrees of freedom in the vector multiplet are represented by the chiral superfields $\Lambda ( \t, z )$ and 
$\Lambda^\dagger ( \t, z )$ as follows: 
\bea
\label{eq:Wess-Zumino-eq1}
e^{2V} 
& \equiv & 
e^{+ i \Lambda^\dagger}e^{2V |_{{\rm WZ}}} e^{ - i \Lambda}\ , \nn 
V |_{{\rm WZ}} ( \t, z )
& = & 
- \theta\sigma^m {\bar \theta} v_m ( \t, x )        
 +
i \theta^2 {\bar \theta}  {\bar \lambda} ( \t, x )         
- i {\bar \theta}^2 \theta \lambda ( \t, x ) 
+ \displaystyle{\frac{1}{2}}\theta^2 {\bar \theta}^2  D ( \t, x ) \ . 
\ena
We call this redefinition of the vector superfield the \lq\lq almost Wess-Zumino gauge\rq\rq. 
As shown in Appendix B, it follows from (\ref{eq:4DLangevin-eq8}) and (\ref{eq:Wess-Zumino-eq1}) that 
the superfield Langevin equation in the \lq\lq almost Wess-Zumino gauge\rq\rq is given by 
\bea
\label{eq:4DLangevin-Wess-Zumino-eq1}
\Delta V|_{\rm WZ}   
& + & 
\displaystyle{\frac{L_V|_{\rm WZ}}{(e^{2L_V|_{\rm WZ}} - 1 )}} 
\Delta \Theta^\dagger 
+ \displaystyle{\frac{L_V|_{\rm WZ}}{( 1 - e^{- 2L_V|_{\rm WZ}} )}} 
 \Delta \Theta                     \nn 
& = & 
 \Delta \t \displaystyle{\frac{1}{4g^2}}  \Big\{ 
\displaystyle{\frac{L_V|_{\rm WZ}}{( 1 - e^{- 2L_V|_{\rm WZ}} )}}  {\cal D}{\cal W}|_{\rm WZ}                                \nn 
& {} & \qquad 
+ \displaystyle{\frac{L_V|_{\rm WZ}}{(e^{2L_V|_{\rm WZ}} - 1 )}}  {\bar {\cal D}}{\bar {\cal W}}|_{\rm WZ} 
\Big\} + \Delta_w \Xi|_{\rm WZ}                                                             \ ,  \nn 
\Delta_w \Xi|_{\rm WZ} 
& \equiv & 
\displaystyle{\frac{L_V|_{\rm WZ}}{(e^{2L_V|_{\rm WZ}} - 1 )}}  \Delta w|_{\rm WZ}     \ .
\ena
Here, the suffix \lq\lq $|_{\rm WZ}$\rq\rq\ indicates that the quantity is evaluated with respect to the vector superfield in the Wess-Zumino gauge, $V|_{\rm WZ} ( \t, z )$.
The time evolution of the auxiliary fields has been collected in the chiral superfields $\Delta \Theta ( \t, z )$ and $\Delta \Theta^\dagger ( \t, z )$: 

\bea
\label{eq:Wess-Zumino-eq2}
\Delta \Theta^\dagger 
& \equiv & 
\displaystyle{\frac{( 1 - e^{- iL_{\Lambda^\dagger}} )}{ L_{\Lambda^\dagger}}}\Delta \Lambda^\dagger        \ , \nn 
\Delta \Theta
& \equiv & 
\displaystyle{\frac{(e^{-i L_\Lambda} - 1 )}{ L_\Lambda}}\Delta \Lambda        \ . 
\ena
$\Delta \Theta$ and $\Delta \Theta^\dagger$ consist of the auxiliary component fields in the vector multiplet and their derivatives with respect to the stochastic time. From the chiral conditions ${\bar D}_{\dot \alpha} \Delta \Theta = D_\alpha \Delta \Theta^\dagger = 0$, 
these collective chiral superfields can be expanded as 
\bea
\label{eq:Wess-Zumino-eq3}
\Delta \Theta ( \t, z ) 
& = & 
\Delta \Theta ( \t, y, \theta )                              \ , \nn 
& \equiv & 
  \Delta {\tilde C} ( \t, y ) + 2i \theta \Delta {\tilde \chi} ( \t, y ) + i \theta^2 ( \Delta {\tilde M} ( \t, y ) + i \Delta {\tilde N} ( \t, y ))                                    \ , \nn
\Delta \Theta^\dagger ( \t, z ) 
& = & 
\Delta \Theta^\dagger ( \t, y^\dagger, {\bar \theta} )       \ , \nn
& \equiv & 
 \Delta {\tilde C} ( \t, y^\dagger ) - 2i {\bar \theta} \Delta {\bar {\tilde \chi}} ( \t, y^\dagger ) - i {\bar \theta}^2 ( \Delta {\tilde M} ( \t, y^\dagger ) - i \Delta {\tilde N} ( \t, y^\dagger ))                            \ , 
\ena
where 
$ y^m = x^m + i\theta\sigma^m {\bar \theta}$ and 
$ {y^m}^\dagger = {x^m}^\dagger - i\theta\sigma^m {\bar \theta}$. 
Equation (\ref{eq:4DLangevin-Wess-Zumino-eq1}) shows that all the auxiliary fields remain, while the time evolution of these auxiliary fields, 
(${\tilde C}$, ${\tilde \chi}$, ${\tilde {\bar \chi}}$, ${\tilde M}$, ${\tilde N}$), is completely determined by the components in the Wess-Zumino gauge, $( v_m, \lambda, {\bar \lambda}, D )$. 

The residual component of the local gauge transformation is now given by $ a (y) $, where $a (x)$ is an arbitrary real function and  
$a (y)^\dagger = a (y^\dagger )$. 
The transformation property for the components in the Wess-Zumino gauge is given by 
$
v_m (x) \rightarrow v_m (x) + {\cal D}_m a (x)   \ , 
\lambda (x) \rightarrow  \lambda (x) + i[ \lambda (x),\ a (x) ] 
$ 
and 
$
D (x) \rightarrow  D (x) + i[ D (x),\ a(x) ]\ .
$

The noise superfield in the Wess-Zumino gauge, 
$\Delta w |_{{\rm WZ}} ( \t, z )$, yields the same correlation as that satisfied by  $\Delta w ( \t, z )$. 
As a consequence, the correlation between $\Delta w |_{{\rm WZ}}$ and $\Delta w |_{{\rm WZ}}^\dagger$ is reduced from (\ref{eq:noise-correlation-eq1}) as  
\bea
\label{eq:Wess-Zumino-nise-eq1}
\langle 
\Delta w |_{{\rm WZ}} ( \t, z ) \Delta w |_{{\rm WZ}}^\dagger ( \t, z' ) 
\rangle_{\Delta w_\t} 
= 2{\rm k} \Delta \t \Big( \delta_{il}\delta_{kj} - \displaystyle{\frac{1}{N}} \delta_{ij}\delta_{kl} \Big) \delta ( z-z' )        \ .
\ena
The hermiticity condition is also given by  
\bea
\label{eq:Wess-Zumino-nise-eq2}
\Delta w |_{{\rm WZ}}^\dagger ( \t, z ) = e^{-{2L_V} |_{{\rm WZ}}} \Delta w |_{{\rm WZ}} ( \t, z )    \ .
\ena

The decomposition of the superfield Langevin equation in the \lq\lq almost Wess-Zumino gauge\rq\rq\ is given in Appendix C. Here we only give the result after eliminating the auxiliary fields. For the components in the Wess-Zumino gauge, $( v_m, \lambda_\alpha, {\bar \lambda}^{\dot \alpha}, D )$, the Langevin equations are given by
\bea
\label{eq:4DLangevin-Wess-Zumino-eq2}
\Delta v_m
& = & 
- \displaystyle{\frac{\Delta\t}{2}} \displaystyle{\frac{1}{g^2}} ( 
{\cal D}^n v_{mn} + \lambda \sigma_m {\bar \lambda} + 
{\bar \lambda}{\bar \sigma}_m \lambda )  +  \Delta W_{v_m}    \ , \nn 
\Delta \lambda 
& = & 
- \displaystyle{\frac{\Delta\t}{2}} \displaystyle{\frac{1}{g^2}} 
\sigma^m {\bar \sigma}^n {\cal D}_m {\cal D}_n \lambda 
+ \Delta W_{\lambda} 
+ \displaystyle{\frac{1}{4}} \sigma^m {\cal D}_m \Delta w|_{\bar \theta}                   \ , \nn 
\Delta {\bar \lambda} 
& = & 
- \displaystyle{\frac{\Delta\t}{2}} \displaystyle{\frac{1}{g^2}} 
{\bar \sigma}^m \sigma^n {\cal D}_m {\cal D}_n {\bar \lambda} 
+ \Delta W_{\bar \lambda} 
- \displaystyle{\frac{1}{4}} {\bar \sigma}^m {\cal D}_m \Delta w |_\theta              \ , \nn 
\Delta D 
& = & 
\displaystyle{\frac{\Delta\t}{2}}  \displaystyle{\frac{1}{ g^2}}{\cal D}^m {\cal D}_m D + 
\Delta W_D 
- \displaystyle{\frac{1}{4}} {\cal D}^m {\cal D}_m \Delta w |    \ .
\ena
Here, we have introduced the collective noise fields $\Delta W_{v_m}$, $\Delta W_{\lambda}$, $\Delta W_{\bar \lambda}$ and $\Delta W_D$, which consist of the components of the noise superfields. These collective noise fields are defined by
\bea
\label{eq:Wess-Zumino-noise-eq3}
\Delta W_{v_m} 
& \equiv & 
- \displaystyle{\frac{1}{4}}\Delta w |_{\theta \sigma^m {\bar \theta}}    
- \displaystyle{\frac{1}{4}}\Delta w^\dagger |_{\theta \sigma^m {\bar \theta}}       \ , \nn     
\Delta W_{\lambda}  
& \equiv & 
\displaystyle{\frac{i}{4}} \Delta w |_{\theta {\bar \theta}^2} 
+ \displaystyle{\frac{i}{4}} \Delta w^\dagger |_{\theta {\bar \theta}^2}                      \ ,\nn 
\Delta W_{\bar \lambda}  
& \equiv & 
- \displaystyle{\frac{i}{4}} \Delta w |_{\theta^2 {\bar \theta}} 
- \displaystyle{\frac{i}{4}} \Delta w^\dagger |_{\theta^2 {\bar \theta}}                \ ,\nn 
\Delta W_D 
& \equiv & 
\displaystyle{\frac{1}{2}} \Delta w |_{\theta^2 {\bar \theta}^2} 
+ \displaystyle{\frac{1}{2}} \Delta w^\dagger |_{\theta^2 {\bar \theta}^2}            \ .
\ena

In order to confirm that these Langevin equations reproduce the action for SYM$_4$ in the Wess-Zumino gauge (\ref{eq:4Daction}) in the equilibrium limit, we clarify the kernel structure of these Langevin equations. 
The correlations for the noise component fields are given by 
\bea
\label{eq:Wess-Zumino-noise-eq4}
& {} & 
\langle  
( \Delta W_{v_m} )_x^a ( \Delta W_{v_n} )_y^b 
\rangle_{\Delta w_\t}                     \nn 
& {} & \qquad 
 =  
 - \Delta \t \delta^{ab} \eta_{mn}  \delta^4 ( x-y )                 \ , \nn
& {} & 
\langle  
( \Delta W_{\lambda} 
+ \displaystyle{\frac{1}{4}} \sigma^m {\cal D}_m \Delta w|_{\bar \theta} )_{x \alpha}^a 
( \Delta W_{\bar \lambda} 
- \displaystyle{\frac{1}{4}} {\bar \sigma}^m {\cal D}_m \Delta w |_\theta )_y^{b {\dot \beta}}
\rangle_{\Delta w_\t}                     \nn 
& {} & \qquad 
 = - i \Delta\t ( \sigma^m \epsilon )_\alpha^{\ {\dot \beta}} 
 ({\cal D}_m)_x^{\ ab} \delta^4 ( x-y )                   \ , \nn
& {} & 
\langle  
( \Delta W_D 
- \displaystyle{\frac{1}{4}} {\cal D}^m {\cal D}_m \Delta w | )_x 
( \Delta W_D 
- \displaystyle{\frac{1}{4}} {\cal D}^m {\cal D}_m \Delta w | )_y 
\rangle_{\Delta w_\t}                     \nn 
& {} & \qquad 
 =  
 - \Delta \t \delta^{ab}  
  {\cal D}^m {\cal D}_m \delta^4 ( x-y )                   \ .
\ena
This result shows that the kernels for the Langevin equations for $v_m$, $( \lambda, {\bar \lambda} )$ and  $D$ are given by 
$- \displaystyle{\frac{1}{2}}$, 
$- \displaystyle{\frac{i}{2}} ( \sigma^m \epsilon )_\alpha^{\ {\dot \beta}} 
 ({\cal D}_m)_x^{\ ab}$ 
 and 
$- \displaystyle{\frac{1}{2}}   
  {\cal D}^m {\cal D}_m $, 
respectively. 
The appearance of the kernels is inevitable, because the canonical dimension of the stochastic time is $[ \Delta \t ] = 2-d $ (where $d$=4 in this case). Without them, we would have to introduce some dimensional parameters artificially into the Langevin equations for the fermion fields and the auxiliary fields on dimensional grounds. However, these parameters would break the supersymmetry at finite stochastic time, because the time development of each component field would be scaled in a different way. These kernels in the component expressions of the superfield Langevin equation and the noise correlations  do not affect the equilibrium distribution. 

We now add a comment on the numerical factor $-{1\over 2}$ of the kernels. This negative factor is absorbed through the Wick rotation to Euclidean space. The $U(1)$ limit of the superfield kernel contributes ${1\over 4}$. The remaining factor $2$ comes from the normalization of the Langevin equation. In fact, even in the $U(1)$ case, we can start from the superfield Langevin equation 
$
\Delta V 
= -\Delta \t {1\over 2}{\delta S\over \delta V} + \Delta w 
$
 and the noise correlation 
$
\langle 
\Delta w^a \Delta {w^b}' 
\rangle 
= \Delta \t \delta^{ab} \delta (z-z') ,
$
in order to obtain the component Langevin equations without any numerical kernel factors, employing the notation of Ref.~\citen{WB}. Our results coincide with those obtaind in Ref.~\citen{BGZ} in the $U(1)$ limit.

In the last part of this section, we briefly illustrate the stochastic gauge fixing procedure\cite{Zwanziger} for SYM$_4$ in the superfield formalism. The gauge fixing procedure is not necessary in the construction of the collective field theory of Wilson loops. However, in the perturbative analysis with gauge variant quantities, there are two important issues in the context of SQM. One is to show that the SQM approach simulates the contribution of the Faddeev-Popov ghost in the path-integral approach. The other is the perturbative renormalizability in SQM. 
For the Yang-Mills case, the stochastic gauge fixing procedure is equivalent to the Faddeev-Popov prescription in the path-integral method.\cite{Zwanziger,BZ,HHS} The perturbative renormalizability of Yang-Mills theory in SQM is shown by introducing the stochastic action principle.\cite{Gozzi,NY,CH} In particular, the BRST symmetry of the stochastic action enables us to apply a standard proof in terms of the Ward-Takahashi identites for the 1-P-I effective stochastic action.\cite{ZZ} 
For SYM$_4$ in the superfield formalism, the stochastic gauge fixing procedure introduces additional auxiliary fields, a chiral and an anti-chiral superfield corresponding to the gauge degrees of freedom, through the local gauge transformation.\cite{Nakazawa3} Let us consider the stochastic-time dependent (inverse) local gauge transformation 
$
e^{2V} 
\rightarrow 
e^{i \Lambda^\dagger} e^{2V} e^{-i \Lambda}
$, where 
$D_\a \Lambda^\dagger = 0$ and ${\bar D}_{\dot \a} \Lambda = 0$, and we define the auxiliary chiral and anti-chiral superfields in a manner similar to that in (\ref{eq:Wess-Zumino-eq2}) for the \lq\lq almost Wess-Zumino gauge\rq\rq : 
\bea
\label{eq:stochastic-gauge-fixing-eq1}
\Phi^\dagger 
& \equiv & 
i 
\displaystyle{\frac{e^{ -i L_{\Lambda^\dagger}} - 1}{L_{\Lambda^\dagger}}} 
\displaystyle{\frac{\Delta \Lambda^\dagger}{\Delta \t}}   \ , \nn
\Phi 
& \equiv & 
i 
\displaystyle{\frac{e^{ -i L_{\Lambda}} -1 }{L_{\Lambda}}} 
\displaystyle{\frac{\Delta \Lambda}{\Delta \t}}   \ , 
\ena
where $D\Phi^\dagger = {\bar D}\Phi = 0$. 
Then we obtain 
\bea
\label{eq:stochastic-gauge-fixing-eq2}
\Delta V  
& + & 
i \displaystyle{\frac{L_V}{(e^{2L_V} - 1 )}} \Phi^\dagger
\Delta \t
- i \displaystyle{\frac{L_V}{( 1 - e^{- 2L_V} )}} \Phi
 \Delta \t                    \nn 
& = & 
 \Delta \t \displaystyle{\frac{1}{4g^2}}  \Big\{ 
\displaystyle{\frac{L_V}{( 1 - e^{- 2L_V} )}}  {\cal D}{\cal W}     
+ \displaystyle{\frac{L_V}{(e^{2L_V} - 1 )}}  {\bar {\cal D}}{\bar {\cal W}}
\Big\} + \Delta_w \Xi                                \ ,  \nn 
\Delta_w \Xi 
& \equiv & 
\displaystyle{\frac{L_V}{(e^{2L_V} - 1 )}}  \Delta w     \ .
\ena
It is obvious that this Langevin equation is covariant under the arbitrary (stochastic-time dependent) local gauge transformation 
$
e^{2V} 
\rightarrow 
e^{-i \Sigma^\dagger} e^{2V} e^{i \Sigma}
$ 
and 
$
e^{i \Lambda} 
\rightarrow 
e^{i \Lambda} e^{i \Sigma } 
$, 
where the superfields $\Sigma$ and $\Sigma^\dagger$ are stochastic-time dependent chiral and anti-chiral, respectively.
The transformation of $\Lambda$ and $\Lambda^\dagger$ induces a transformation of $\Phi$ and $\Phi^\dagger$: 
\bea
\label{eq:stochastic-gauge-fixing-eq3}
\Phi^\dagger 
& \rightarrow &
e^{-i L_{\Sigma^\dagger}} \Phi^\dagger 
+ 
i \displaystyle{\frac{( e^{ -i L_{\Sigma^\dagger}} - 1 )}{L_{\Sigma^\dagger}}} 
\displaystyle{\frac{\Delta \Sigma^\dagger}{\Delta \t}}        \ , \nn
\Phi 
& \rightarrow &
e^{-i L_\Sigma} \Phi 
+ 
i \displaystyle{\frac{(e^{-i L_\Sigma} - 1)}  {L_\Sigma}} 
\displaystyle{\frac{\Delta \Sigma}{\Delta \t}}        \ .
\ena
The stochastic gauge fixing term does not contribute to the expectation value of gauge invariant observables in the equilibrium limit. This can be shown by using the corresponding time evolution equation for observables in which the additional term appears as the generator of the infinitesimal local gauge transformation, 
\bea
\label{eq:stochastic-gauge-fixing-eq4}
G( \Phi, \Phi^\dagger ) 
& = & 
\displaystyle{\frac{i}{\rm k}}
\int \!\!\!dz {\rm Tr} \Big( \Phi^\dagger (z) - e^{2L_V} \Phi (z) \Big) {\hat {\cal E}} (z)  ,
\ena
which forms the algebra  
\bea
\Big[ 
 G( \Phi_1, \Phi_1^\dagger ),\ G( \Phi_2, \Phi_2^\dagger ) \Big] 
 = -i G\Big( [ \Phi_1, \Phi_2 ],\ [ \Phi_1^\dagger, \Phi_2^\dagger ] \Big)  
     \ . 
\ena
In terms of the Langevin equation (\ref{eq:stochastic-gauge-fixing-eq2}), the stochastic gauge fixing procedure is defined by specifying $\Phi$ and $\Phi^\dagger$, which play the role of the Zwanziger gauge fixing functions, as functions of the vector superfield $V$. For example, we can choose $\Phi$ and $\Phi^\dagger$ as $i \a {\bar D}^2 D^2 V$ and $-i \a D^2 {\bar D}^2 V$, respectively, with $\a$ the gauge parameter. These gauge fixing functions are almost uniquely determined by the condition ${\bar D}\Phi = D \Phi^\dagger = 0$ and dimensional arguments. The gauge fixed superfield Langevin equation reproduces the well-known superpropagator with a gauge parameter\cite{FP} defined in the path-integral approach in the equilibrium limit. In fact, it can be directly confirmed that this stochastic gauge fixing term in the superfield formalism includes one that appears in the Yang-Mills case\cite{Zwanziger} in the component expansion. 
For this type of stochastic gauge fixing, it is also possible to prove that the Faddeev-Popov probability distribution\cite{FP} is reproduced in the equilibrium limit.\cite{Nakazawa3} This is a simple extension of the Yang-Mills case\cite{BZ,HHS} to SYM$_4$. 

Another important issue is the perturbative renormalizability, including the non-renormalization theorem, which is a particular feature of supersymmetric theories, in the context of the SQM approach. The renormalizability as well as the non-renormalization theorem are demonstrated by introducing the BRST invariant structure in terms of the stochastic action principle. Since (\ref{eq:stochastic-gauge-fixing-eq3}) defines a \lq\lq 5-th\rq\rq dimensional local gauge symmetry, it is straightforward to introduce a 5-dimensional BRST symmetry, as is well known in the Yang-Mills case.\cite{CH,KOT,Nakazawa2} Then, a consistent truncation of the 5-dimensional BRST symmetry yields a BRST symmetry in the extended phase space, following the general method for the construction of BRST invariant stochastic action for first-class constraint systems.\cite{Nakazawa2,ZZ} The BRST symmetry of the stochastic action yields the Ward-Takahashi identities, which, with the power counting argument, ensure the renormalizability of the BRST invariant stochastic action. The stochastic action principle is also useful for an explicit perturbative calculation. In order to preserve both the local gauge symmetry and the global supersymmetry, we employ regularization by dimensional reduction.\cite{Siegel-CJN} It is possible to apply the background field method in SQM\cite{Okano-MOSZ} to SYM$_4$ in terms of the stochastic action principle. The perturbative analysis is consistent with the path-integral approach for SYM$_4$\cite{GSR} at the one-loop level. The details of this point are reported in Ref.\cite{Nakazawa4}

Here we comment on the difference between the stochastic gauge fixing procedure and the almost Wess-Zumino gauge. In the stochastic gauge fixing procedure, we choose the Zwanziger gauge fixing functions, $\Phi$ and $\Phi^\dagger$, as functions of the vector superfield, $V$. This breaks the extended local gauge covariance of the Langevin equation, while it preserves the global supersymmetry manifestly in the superfield formalism. The time development of the local gauge invariant observables expressed with the superfield does not depend on the gauge fixing functions. On the other hand, in view of the stochastic gauge fixing procedure, we identify the Zwanziger gauge fixing functions with the derivatives of the auxiliary component fields in the vector multiplet with respect to the stochastic time in the almost Wess-Zumino gauge, as is clear from (\ref{eq:Wess-Zumino-eq2}) and (\ref{eq:stochastic-gauge-fixing-eq1}). In this case, since the vector superfield is restricted to the Wess-Zumino gauge, $V|_{\rm WZ}$, we need to perform an extended local gauge transformation to define the supersymmetry transformation within the Wess-Zumino gauge.   
We note that although the time development of the auxiliary fields remains, the time development of local  gauge invariant observables does not depend on the auxiliary fields. The reason for this is the same as in the case of the stochastic gauge fixing procedure.

%

\section{Langevin equations and the Fokker-Planck equation for the physical component fields in Euclidean SYM$_4$}

For an extension to SYM$_{10}$, we reconstruct the Langevin equation for SYM$_4$ with a 4-dimensional Majorana spinor by starting from the action of SYM$_4$ in Minkowski space-time and explain the Euclidean prescription for the Majorana fermion. SYM$_4$ in Minkowski space-time is defined by 
\bea
\label{eq:4Daction-component1}
S 
= 
\int\!\!\!d^4\!\! x \displaystyle{\frac{1}{g^2}} {\rm Tr}\big( 
- {1\over4} v_{mn}v^{mn} + {i\over 2} {\bar \psi}\caldslash \psi
\big)         \ .  
\ena
Here, $\psi$ is a Majorana spinor: 
$\psi^t \equiv ( (\lambda_\alpha )^t , ({\bar \lambda}^{\dot \alpha} )^t )$. $\gamma^m$ are 4-dimensional $\gamma$-matrices in a Weyl representation that satisfy 
$\big\{ \gamma^m, \gamma^n \big\} = -2 \eta^{mn}$, with 
$\eta^{mn}= (-,+,+,+)$. 
The charge conjugation matrix ${\cal C}$ satisfies 
$
( \gamma^m )^{t} 
= - {\cal C}^{-1}\gamma^m {\cal C}    \ , 
{\cal C}^{t} 
= - {\cal C}    . 
$
The Majorana fermion satisfies the relation 
$
\psi^c \equiv {\cal C} {\bar \psi}^t = \psi
$ .
The model possesses local gauge symmetry and  ${\cal N}$=1 supersymmetry: 
\bea
\label{eq:4Dsupersymmetry-component}
\d A_\mu
& = & 
i{\bar \epsilon}\gamma_m \psi \ ,  \nn
\d \psi 
& = & 
-{1\over 2}v^{mn}\gamma_{mn}\epsilon  \ . 
\ena

We fix the Wick rotation prescription for the Majorana fermion, following Nicolai,\cite{Nicolai} though the Majorana spinor does not exist in the Euclidean space, that keeps the explicit connection of the Euclidean theory to the Minkowski one. We introduce the independent Majorana spinors $\psi$ and ${\bar \psi}$ and impose the constraint 
\bea
\label{eq:4DMajorana-condition}
{\cal C} {\bar \psi}^t = \psi    \ .
\ena
We then perform the Wick rotation, 
$
x^0 = -i x^4  ,\ \gamma^0 =-i \gamma^4 \ . 
$ 
The expression of the Euclidean action is given by 
$
iS \equiv - S_E , 
$
 where
\bea
\label{eq:4Daction-component2}
S_E 
= 
\int\!\!\!d^4\!\! x \displaystyle{\frac{1}{g^2}} {\rm Tr}\big( 
{1\over4} v_{mn}^2 - {i\over 2} {\bar \psi}\caldslash \psi
\big)         \ .   
\ena
The Euclidean supersymmetry transformation is the same as that in Minkowski space-time. To prove the supersymmetry of the Euclidean action, it is sufficient to note that the \lq\lq charge conjugation \rq\rq matrix satisfies a relation such as 
$
\gamma_m^{t} 
= - {\cal C}^{-1}\gamma_m {\cal C} , 
$ 
with 
$ 
m=1,2,3,4 \ .
$
%


The time evolution of the component fields in (\ref{eq:4DLangevin-Wess-Zumino-eq2}), for $v_m ( \t+\Delta \t ) \equiv v_m ( \t ) + \Delta v_m ( \t )$, $\psi ( \t+\Delta \t ) \equiv \psi ( \t ) + \Delta \psi ( \t )$, is described by the following Langevin equation with the Majorana fermion. It reproduces the probability distribution $e^{-S_E}$ with $S_E$ in (\ref{eq:4Daction-component2}) for $\t \rightarrow \infty$: 
\bea
\label{eq:4DLangevin-component}
{\Delta}v_m (\t)
& = & 
-\Delta\t \displaystyle{\frac{1}{g^2}} \big( 
{\cal D}_n v_{mn} - {\bar \psi}\gamma_m \psi 
\big) (\t) + \Delta\xi_m (\t)          \ , \nn
{\Delta}\psi (\t)
& = & 
- \Delta\t \displaystyle{\frac{1}{g^2}} \caldslash\ ^2 \psi (\t) 
+ \Delta \chi ( \t )          \ , \nn
\Delta \chi ( \t )
& = & 
\Delta\xi (\t) + i\caldslash \Delta\eta (\t)  \ .
\ena
Since the Majorana fermion fields, $\psi$ and ${\bar \psi}$, are not independent, the Langevin equation for ${\bar \psi}$ is given by \lq\lq charge conjugation\rq\rq,  
$
\Delta{\bar \psi} (\t)
 = 
 -\Delta\t {\bar \psi} \overleftarrow{\caldslash}\ ^2(\t) 
+ {\overline {\Delta\chi}} (\t) 
        \ 
$
with
$
{\overline {\Delta\chi}} (\t)         
 =  
 {\overline {\Delta\xi}} (\t) - i {\overline {\Delta\eta}} \overleftarrow{\caldslash}\ (\t)   \ ,
$
where we write 
$
{\bar \psi}\overleftarrow{{\cal D}_m} 
\equiv {\cal D}_m{\bar \psi}  . 
$
We have introduced the noise variables $\Delta\xi_\mu$, $\Delta\xi$ and $\Delta\eta$, which are a 4-dimensional vector and two Majorana spinors. All these noise variables are $SU(N)$ algebra-valued. Their correlations are defined by 
\bea
\label{eq:4Dnoise-component}  
\langle 
\Delta\xi^a_m(\t,x) \Delta\xi^b_n(\t,y) 
\rangle 
& = & 
2\Delta\t \d^{ab}\d_{mn}  \d^4(x-y)     \ , \nn  
\langle 
\Delta\xi^a_\a(\t,x) {\overline {\Delta\eta}}^b_\b(\t,y) 
\rangle  
& = & 
- \langle 
{\overline {\Delta\eta}}^b_\b(\t,y) \Delta\xi^a_\a(\t,x) 
\rangle     
= 
\Delta\t \d^{ab}\d_{\a\b} \d^4(x-y)     .  
\ena
This yields the correlation for the collective noise $\Delta \chi$ as 
\bea
\label{eq:4Dcollective-correlation}
\langle 
\Delta\chi^a_\a(\t,x) \Delta\chi^b_\b(\t,y) 
\rangle  
= 
2\Delta\t ( \Gamma_\mu{\cal C} 
)_{\a\b} (-iD_\mu^{ab})_x  \d^4(x-y)          .  
\ena
%


The Fokker-Planck equation for the component fields is also derived from (\ref{eq:4DLangevin-component}). The time evolution of an observable ${\cal O}( v_m, \psi_\alpha )$ is given by 
$
\Delta {\cal O} 
= 
- \Delta\t {\cal H}_{\rm FP} {\cal O}     ,       
$
with
\bea
\label{eq:4DFP-hamiltonian}
{\cal H}_{\rm FP} 
& = & 
\int\!\!\! d^4x \Big\{ \displaystyle{\frac{1}{g^2}}\Big( 
 {\cal D}_n v_{mn} - {\bar \psi}\gamma_m \psi \Big)^a (x) - \displaystyle{\frac{\d\quad }{\d v^a_m(x)}} \Big\} \displaystyle{\frac{\d\quad }{\d v^a_m(x)}}             \nn
& + & 
\int\!\!\! d^4x ( 
\gamma_m{\cal C} 
) _{\a\b} (-i{\cal D}_m)^{ab}_x  \Big\{  \displaystyle{\frac{1}{g^2}}\Big( 
 i {\cal C}^{-1} \caldslash\ \psi (x) \Big) ^b_\b  
- \displaystyle{\frac{\d\quad }{\d \psi^b_\b(x)}}  \Big\} \displaystyle{\frac{\d\quad }{\d \psi^a_\a(x)}}                   .
\ena
Here, we have used the left derivative for the fermionic variables. 
We also obtain the following Fokker-Planck equation for the probability distribution: 
\bea
\label{eq:4DFokker-Planck-eq-component}
\displaystyle{\frac{\pa P}{\pa \t}}
& = &  
\int\!\!\! d^4x\displaystyle{\frac{\d }{\d v^a_m}}\Big\{ \Big( \displaystyle{\frac{\d S_E}{\d v^a_m}} 
+ \displaystyle{\frac{\d }{\d v^a_m}} \Big) P \Big\}   \nn
& + & 
\int\!\!\! d^4x\displaystyle{\frac{\d }{\d \psi^a_\a}}\big( 
(-i\caldslash \ ) {\cal C} 
\big)^{ab}_{\a\b} \Big\{ \Big( \displaystyle{\frac{\d S_E}{\d \psi^b_\b}} + \displaystyle{\frac{\d }{\d \psi^b_\b}} \Big) P \Big\}  \ . 
\ena
Hence, we obtain the action (\ref{eq:4Daction-component2}) in the equilibrium limit,  
$ 
\lim_{\t \rightarrow \infty} P = e^{- S_E} 
$ . 


\section{SYM$_{10}$ in SQM}
Now, we consider the ten-dimensional case.  We begin with the Euclidean action of $SU(N)$ SYM$_{10}$, 
\bea
\label{eq:10Daction}
S 
= 
\int\!\!\!d^{10}\!\! x \displaystyle{\frac{1}{g^2}} {\rm Tr}\big( 
{1\over4} F_{\mu\nu}^2 -{i\over 2} {\bar \Psi}\dslash \Psi
\big)         \ ,   
\ena
where $A_\mu$ and $\Psi$, which are a vector and a Majorana-Weyl spinor in ten dimensions, are $SU(N)$ algebra-valued. 
$
F_{\mu\nu} = \pa_\mu A_\nu - \pa_\nu A_\mu - i\Big[ 
A_\mu, A_\nu \Big] 
$
and 
$
D_\mu \Psi = \pa_\mu \Psi - i \Big[ 
A_\mu, \Psi \Big] 
$. 
The 
$\Gamma_\mu$ are 10-dimensional $\gamma$-matrices which satisfy 
$\big\{ \Gamma_\mu, \Gamma_\nu \big\} = -2\delta_{\mu\nu}$. We use the convention 
$
\Gamma_j = i\sigma_1\otimes\gamma_j\ $ for $j=1,...,8, \Gamma_9 = i\sigma_3\otimes {\bf 1}_{16} 
$
and 
$
\Gamma_{10} = i\sigma_2\otimes{\bf 1}_{16} 
$. With real symmetric $\gamma_i$ that satisfy $\{ \gamma_i, \gamma_j \} = 2\d_{ij}$, all the $\Gamma_\mu$ are anti-hermitian, and 
$
\Gamma_{11} = i\Gamma_1 ... \Gamma_{10} \ 
$
is real symmetric.
For later convenience, we fix the chirality of the spinor as $\Gamma_{11} \Psi = \Psi$. Here, ${\bar \Psi}$ is defined by \lq\lq charge conjugation\rq\rq: 
$
{\bar \Psi} = - \Psi^t {\cal C} .
$
The charge conjugation matrix ${\cal C}$ satisfies the relation 
$
\Gamma_\mu 
= 
- {\cal C}^{-1}\Gamma_\mu^{t} {\cal C}    \ , 
$ 
where 
$ 
{\cal C}^{t} 
= 
- {\cal C}    \ ,
$
which ensures the relations necessary to prove the supersymmetry of the Euclidean action, such as 
${\bar \epsilon} \Gamma_{\mu_1}\cdots \Gamma_{\mu_M}\Psi 
= 
(-)^{M}{\bar \Psi} \Gamma_{\mu_M}\cdots \Gamma_{\mu_1}\epsilon \ .$ 
The model possesses local gauge symmetry and ${\cal N}$=1 supersymmetry: 
$
\d A_\mu
= 
i{\bar \epsilon}\Gamma_\mu \Psi \ , 
\d \Psi 
= 
-{1\over 2}F_{\mu\nu}\Gamma_{\mu\nu}\epsilon  \ . 
$

The Langevin equations for SYM$_{10}$ are given by 
\bea
\label{eq:10DLangevin}
{\Delta}A_\mu (\t)
& = & 
-\Delta\t \displaystyle{\frac{1}{g^2}} \big( 
D_\nu F_{\mu\nu} + {\bar \Psi}\Gamma_\mu \Psi 
\big) (\t) + \Delta\xi_\mu(\t)          \ , \nn
{\Delta}\Psi (\t)
& = & 
- \Delta\t \displaystyle{\frac{1}{g^2}} \dslash\ ^2 \Psi (\t) 
+ \Delta\xi (\t) + i\dslash \Delta\eta (\t)          \  . 
\ena
Here, we have introduced the noise variables $\Delta\xi_\mu$, $\Delta\xi$ and $\Delta\eta$, a 10-dimensional vector and two 10-dimensional Majorana-Weyl spinors. All these noise variables are $SU(N)$ algebra-valued. The main difference between these Langevin equations and those for SYM$_4$ appears in the following noise correlations, which include the chiral projection corresponding to the Majorana-Weyl condition on the noise variables: 
\bea
\label{eq:10Dnoise}  
\langle 
\Delta\xi^a_\mu(\t,x) \Delta\xi^b_\nu(\t,y) 
\rangle  
& = & 
2\Delta\t \d^{ab}\d_{\mu\nu}  \d^{10}(x-y)     \ , \nn  
\langle 
\Delta\xi^a_\a(\t,x) {\overline {\Delta\eta}}^b_\b(\t,y) 
\rangle  
& = & 
- \langle 
{\overline {\Delta\eta}}^b_\b(\t,y) \Delta\xi^a_\a(\t,x) 
\rangle     \nn  
& = & 
\Delta\t \d^{ab}{1\over 2}(1+\Gamma_{11})_{\a\b}  \d^{10}(x-y)     \ .  
\ena

Here we note the chirality assignment of the fermionic (Majorana-Weyl) noise variables. The time development must preserve the chirality of the Majorana-Weyl fermion, while the operator $\dslash\ $ flips it. Thus, we define the chirality of the two independent Majorana noises, 
$\Gamma_{11}\Delta \xi = \Delta \xi$ and 
$\Gamma_{11}\Delta \eta = - \Delta \eta$. By the definition $\Delta\eta\equiv {\cal C}^{-1}{\overline {\Delta\eta}}^t$,  (\ref{eq:10Dnoise}) yields 
\bea
\label{eq:10Dnoise2}
\langle 
\Delta\xi^a_\a(\t) \Delta\eta^b_\b(\t) 
\rangle  
= 
- \Delta\t \d^{ab}\Big( {1\over 2}(1+\Gamma_{11}){\cal C}\ ^{-1} \Big)_{\a\b}  \d^{10}(x-y)       \ .  
\ena
By introducing a collective fermionic (Majorana-Weyl) noise variable 
$
\Delta\chi , 
$
the correlation can be written in the compact form  
\bea
\label{eq:10Dcollective-correlation} 
\Delta\chi 
& \equiv & 
 \Delta\xi + i\dslash \Delta\eta   \ ,  \nn
\langle 
\Delta\chi^a_\a(\t,x) \Delta\chi^b_\b(\t,y) 
\rangle  
& = & 
2\Delta\t\Big( {1\over 2}(1+\Gamma_{11})\Gamma_\mu{\cal C}\ ^{-1} 
\Big)_{\a\b} (-iD_\mu^{ab})_x  \d^{10}(x-y)          \ .  
\ena
For the Majorana-Weyl fermion, $\Psi$ and ${\bar \Psi}$ are not independent of each other, and we have 
$
\Delta{\bar \Psi} (\t)
 = 
 -\Delta\t {\bar \Psi} \overleftarrow{\dslash}\ ^2(\t) 
+ {\overline {\Delta\chi}} (\t) 
        \ , 
$
where 
$
{\overline {\Delta\chi}} (\t)         
 =  
 {\overline {\Delta\xi}} (\t) 
 - i {\overline {\Delta\eta}} \overleftarrow{\dslash}\ (\t)   \ .
$
%

The time development of an observable ${\cal O}(A_\mu, \Psi)$ is also defined by 
$
\Delta {\cal O} 
= 
- \Delta\t {\cal H}_{\rm FP} {\cal O}    \      
$, 
with the Fokker-Planck Hamiltonian 
\bea
\label{eq:10DFokker-Planck-hamiltonian}
& {} & {\cal H}_{\rm HP}    \nn
& = & 
\int\!\!\! d^{10}x \Big\{ \displaystyle{\frac{1}{g^2}} \Big( 
 D_\nu F_{\mu\nu} + {\bar \Psi}\Gamma_\mu \Psi \Big)^a (x) - \displaystyle{\frac{\d\quad }{\d A^a_\mu(x)}} \Big\} \displaystyle{\frac{\d\quad }{\d A^a_\mu(x)}}             \nn
& + & 
\int\!\!\! d^{10}x \Big( 
{1\over 2}(1+\Gamma_{11})( -i\dslash \ ){\cal C}\ ^{-1} 
\Big) ^{ab}_{\a\b}  \Big\{  \displaystyle{\frac{1}{g^2}} \Big( 
i {\cal C} \dslash\ \Psi (x) \Big) ^b_\b  
- \displaystyle{\frac{\d\quad }{\d \Psi^b_\b(x)}}  \Big\} \displaystyle{\frac{\d\quad }{\d \Psi^a_\a(x)}}                   .
\ena
The corresponding Fokker-Planck equation is given by 
\bea
\label{eq:10DFokker-Planck-eq1}
\displaystyle{\frac{\pa P}{\pa \t}}
& = &  
\int\!\!\! d^{10}x\displaystyle{\frac{\d }{\d A^a_\mu}}\Big\{ 
\Big( \displaystyle{\frac{\d S}{\d A^a_\mu}} + \displaystyle{\frac{\d }{\d A^a_\mu}} \Big) P \Big\}     \nn
& + & 
\int\!\!\! d^{10}x\displaystyle{\frac{\d }{\d \Psi^a_\a}}\big( 
\displaystyle{\frac{1}{2}}(1+ \Gamma_{11}) (-i\dslash \ ) {\cal C}\ ^{-1}
\big)^{ab}_{\a\b} \Big\{ \Big( \displaystyle{\frac{\d S}{\d \Psi^b_\b}} + \displaystyle{\frac{\d }{\d \Psi^b_\b}} \Big) P \Big\}   \ .
\ena
This formally ensures that the path-integral measure with (\ref{eq:10Daction}) is reproduced in the equilibrium limit,  
$ 
\lim_{\t \rightarrow \infty} P = e^{- S} 
$ .


\section{IIB Matrix Model in SQM}

Now we discuss a zero volume limit of the Langevin equation (\ref{eq:10DLangevin}). One of our main interests is to apply it to the IIB matrix model\cite{IIB} 
and to construct a collective field theory of Wilson loops. 
 For the bosonic part, the construction is illustrated in this context.\cite{EKN}  The IIB matrix model is obtained as a naive zero volume limit of the $SU(N)$ SYM$_{10}$ in (\ref{eq:10Daction}). 
\bea
\label{eq:IIBaction}
S_{{\rm IIB}} 
= 
- \displaystyle{\frac{1}{g^2}}{\rm Tr}\Big( 
{1\over4} [ A_\mu, A_\nu ]^2 +{1\over 2} {\bar \Psi}\Gamma_\mu [ A_\mu, \Psi ] 
\Big)         .  
\ena
This model possesses ${\cal N} = 2$ supersymmetry. One of the ${\cal N} = 2$ results from the zero volume limit of the ${\cal N} = 1$ supersymmetry of SYM$_{10}$, and the other is enhanced in the zero volume limit which is a reminiscence of the Green-Schwarz IIB superstring action:\cite{IIB} 
\bea
\label{eq:IIBsupersymmetry1}
\d_1 A_\mu 
& = & 
i{\bar \epsilon}\Gamma_\mu \Psi \ ,  \nn
\d_1 \Psi    
& = & 
{i\over 2}[ A_\mu, A_\nu ]\Gamma_{\mu\nu}\epsilon  \, \nn
\d_2 A_\mu 
& = & 
0                     \ , \nn
\d_2 \Psi 
& = & 
\lambda{\bf 1}        \ . 
\ena
Here, the transformation parameters, $\epsilon$ and $\lambda$, are Majorana-Weyl spinors in ten dimensions and ${\bf 1}$ is the $N\times N$ unit matrix. In the definition of the IIB matrix model, the matrix variables are $N\times N$ hermitian matrices with non-zero traces. After taking the zero volume limit of $SU(N)$ SYM$_{10}$, the model is extended to be $U(N)$ algebra-valued. Although a precise proof of the finiteness has been given for the $SU(N)$ algebra-valued case,\cite{AW} the trace part is necessary for the realization of the ${\cal N} = 2$ supersymmetry.  

The reduced version of the Langevin equation is defined by the zero volume limit of (\ref{eq:10DLangevin}), with an extension of the fundametal variables and the noise to be $U(N)$ algebra-valued: 
\bea
\label{eq:IIBLangevin}
{\Delta}A_\mu (\t)
& = & 
-\Delta\t \displaystyle{\frac{1}{g^2}} \Big( 
[ A_\nu, [ A_\nu, A_\mu ] ](\t) + {\bar \Psi}\Gamma_\mu \Psi(\t) 
\Big) + \Delta\xi_\mu(\t)          \ , \nn
{\Delta}\Psi (\t)
& = & 
\Delta\t \displaystyle{\frac{1}{g^2}} \Gamma_\mu 
 [ A_\mu, \Gamma_\nu [A_\nu, \Psi ] ](\t) 
+ \Delta\chi (\t)           \ , \nn
\Delta\chi (\t)  
& \equiv & 
\Delta\xi (\t) + \Gamma_\mu[ A_\mu, \Delta\eta ](\t)   \ .
\ena
The correlation of noise variables is also obtained from the zero volume limit of (\ref{eq:10Dnoise}). It is given by 
\bea
\label{eq:IIBnoise}  
\langle 
\Delta\xi_\mu(\t)_{ij} \Delta\xi_\nu(\t)_{kl} 
\rangle  
& = & 
2\Delta\t \d_{il}\d_{kj}\d_{\mu\nu}      \ , \nn  
\langle 
\Delta\xi_\a(\t)_{ij} {\overline {\Delta\eta}}_\b(\t)_{kl} 
\rangle  
& = & 
- \langle 
{\overline {\Delta\eta}}_\b(\t)_{kl} \Delta\xi_\a(\t)_{ij} 
\rangle     
= 
\Delta\t \d_{il}\d_{kj}{1\over 2}(1+\Gamma_{11})_{\a\b}       .  
\ena
This yields 
\bea
\label{eq:IIB-correlation}
\langle 
\Delta\chi_\a(\t)_{ij} \Delta\chi_\b(\t)_{kl} 
\rangle  
= 
2\Delta\t\Big( {1\over 2}(1+\Gamma_{11})\Gamma_\mu{\cal C}\ ^{-1} 
\Big)_{\a\b}\Big\{ \d_{il}(A_\mu)_{kj} - \d_{kj}(A_\mu)_{il} 
  \Big\}        .  
\ena
%


For actual calculations, it is suffcient to consider a generalized Langevin equation for the Wilson loops, 
while, in the field theoretical interpretation of SQM, the Hamiltonian operator is given by the Fokker-Planck Hamiltonian.\cite{EKN} From the zero volume limit of (\ref{eq:10DFokker-Planck-hamiltonian}), we obtain the Fokker-Planck Hamiltonian for the IIB matrix model: 
\bea
\label{eq:IIBF-P hamiltonian}
{\cal H}_{\rm IIB}   
& = & 
\Big\{ \displaystyle{\frac{1}{g^2}} \Big( 
 [ A_\nu, [ A_\nu, A_\mu ] ] + {\bar \Psi}\Gamma_\mu \Psi \Big)_{ij}  - \displaystyle{\frac{\d\quad }{\d ( A_\mu )_{ji}}} \Big\} \displaystyle{\frac{\d\quad }{\d (A_\mu)_{ij} }}             \nn
& - &  
\displaystyle{\frac{1}{g^2}} 
\Big( \Gamma_\mu [ A_\mu, \Gamma_\nu [ A_\nu, \Psi ]]
  \Big)_{\a\ ij}\displaystyle{\frac{\d\quad }{\d ( \Psi_\a )_{ij} }}    \nn
& - & 
\Big( 
{1\over 2}(1+\Gamma_{11})\Gamma_\mu{\cal C}\ ^{-1} 
\Big) _{\a\b} \Big\{ \d_{il} ( A_\mu )_{kj} - \d_{kj} ( A_\mu )_{il}  
\Big\} \displaystyle{\frac{\d^2 \quad }{\d ( \Psi_\b )_{kl} \d ( \Psi_\a )_{ij} }}                  \  . 
\ena 

In principle, we can construct the Lorentz invariant collective field theory of the Wilson loops, 
\bea
W_M 
& \equiv & 
{\rm Tr}\prod_{n=1}^M U_n         \ , \nn
U_n 
& = & 
e^{ i\epsilon ( k_n^\mu A_\mu + {\bar \lambda}_n\Psi ) }   \ , 
\ena
in IIB matrix model from the time evolution described by the Langevin equation. It is also possible by changing the variable to the Wilson loop in the Fokker-Planck Hamiltonian (\ref{eq:IIBF-P hamiltonian}). We now comment on some implications of the Langevin equation (\ref{eq:IIBLangevin}). 
The derivation of the generalized Langevin equation for Wilson loops is straightforward. It is believed to describe the time development of string fields in IIB superstring field theory. However, its continuum limit is non-trivial. Since the expectation value of the generalized Langevin equation yields the Schwinger-Dyson equation (loop equation) in the equilibrium limit, we obtain a linear combination of two Schwinger-Dyson equations, one coming from the Langevin equation for the bosonic matrices and the other from the fermionic matrices, which are essentially the same as those investigated in Ref.~\citen{IIB2} to derive the light-cone IIB string field theory. The difference results from the fact that the Langevin equation for the Majorana-Weyl fermion in (\ref{eq:IIBLangevin}) includes the kernel. This means that the Schwinger-Dyson equation for the fermionic variables is not the insertion of the equation of motion derived from the original IIB matrix model, as it includes an additional covariant derivative. These insertions can be re-expressed in terms of the momenta $k_\mu$ and $\lambda_\a$ and the differential operation with respect to these momenta. In the double scaling limit, this insertion of the additional covariant derivative causes a higher power of $\epsilon$. As a consequence, in the naive continuum limit, some of the contributions from the insertion of the fermionic Langevin equation vanish. Therefore we need either a prescription for the continuum limit in which these contributions remain finite in the continuum limit or to give a physical meaning to these vanishing pieces. There is a known example in which a similar situation exists. In old fashioned matrix models, matrix-vector models describe the non-critical open-closed string theories. In that case, the scaling of the time coordinate along the boundary (open string end points) is different from the scaling on the bulk in the continuum limit.\cite{Kostov,NE} Then, if we ignore the difference in the scaling behavior of the time coordinate, some parts of the string field theory Hamiltonian vanish in the double scaling limit of the matrix-vector model. However, the significant fact is that the higher order of $\epsilon$ (the continuum limit implies $\epsilon \rightarrow 0$) includes the open string interactions and the contributions, which vanish in the naive continuum limit, are necessary to close the algebraic structure of the string field theory Hamiltonian.\cite{Kostov,NE} It is a consequence of the integrability of the non-critical string field theory constraints\cite{Mogami}. It is not obvious at present whether such a mechanism is also effective in the case of the IIB marix model. Futher investigation of the construction of a manifestly Lorentz invariant collective field theory of Wilson loops is under investigation in the context of SQM.


\section{Conclusion}

In this paper, we have applied SQM to SYM both in four and ten dimensions. In the superfield formalism for SYM$_4$, the local gauge symmetry, as well as the global ${\cal N}$ = 1 supersymmetry, is manifestly preserved in the sense of It${\bar {\rm o}}$ calculus. In particular, the Langevin equation and the corresponding Fokker-Planck equation are formulated in a geometrically covariant form in superspace, where a non-trivial path-integral measure is introduced to define the local gauge invariant partition function. This is a manifestation of the covariant nature of It${\bar {\rm o}}$ stochastic calculus. The metric in superspace can be regarded as a kernel of the superfield Langevin equation. In component fields, we obtained kerneled Langevin equations for fermion fields and the auxiliary component fields in the vector multiplet. The kerneled forms of the component Langevin equations are necessary, as demonstrated by dimensional arguments. We have shown that in order to obtain the Langevin equations for the physical component fields in the vector multiplet, it is possible to choose the \lq\lq almost Wess-Zumino gauge\rq\rq, in which the time development of the auxiliary component fields is completely determined by the component fields in the Wess-Zumino gauge. Although all the auxiliary component fields remain, we have obtained the Langevin equations for the component fields in the Wess-Zumino gauge through a consistent truncation of the component Langevin equations. In ten dimensions, though there is no superfield formulation, the structures of the Langevin equations and the Fokker-Planck Hamiltonian are similar to those in the 4-dimensional case, except that, correponding to the Majorana-Weyl condition, the chiral projection is introduced for the fermionic noise correlation. This formulation may be useful for the construction of collective field theories of Wilson loops in SYM both in four and ten dimensions.  We have also derived the Langevin equation and the Fokker-Planck Hamiltonian for the IIB matrix model by taking the zero volume limit of those for SYM$_{10}$. 
Application of our formulation to lower dimensional SYM$_d$ (i.e. $d=3,6$) is straightforward. 

Finally, we add some remarks on the regularization in the SQM approach and the equivalence of SQM and the path-integral method in quantizing SYM. 
In a precise sense, what we have shown is that, apart from the stochastic gauge fixing procedure, which simulates the contribution of the Faddeev-Popov determinant, the path-integral measure $e^{-S}$ for SYM is a stationary solution of the corresponding Fokker-Planck equation. We have assumed the existence of a unique equilibrium with which we expect the stationary solution coincides in the infinite limit of the stochastic time. For a system with a finite number of the degrees of freedom, there is a theorem that asserts the exsistence of the equilibrium limit and its uniqueness under certain conditions for the Fokker-Planck equation and for its stationary solution.\cite{BZ} In order to apply this theorem to the systems discussed in this note, we need a regularization procedure that reduces the number of degrees of freedom of SYM to a finite number by preserving the global supersymmetry as well as the local gauge symmetry. The large N reduction of SYM, realized by taking the naive zero volume limit, can be regarded as such a regularization procedure for finite N. In fact, the finiteness of the partition function and the correlation functions has been proved even for the non-supersymmetric YM case.\cite{AW-KS} In this sense, the reduced model of $SU(N)$ YM theory, for which we have constructed a collective field theory of Wilson loops, may provide an example that has a unique equilibrium limit in the SQM approach and the equilibrium distribution is equivalent to that defined in the path-integral approach. On the other hand, for the supersymmetric case, the fermionic path-integral yields a Pfaffian for the path-integral measure. In general, this Pfaffian is not real. In particular, in the ten-dimensional case, the Pfaffian is complex and it may cause a serious problem in defining the stochastic process. By contrast, in four dimensions, the Pfaffian is positive semi-definite. Therefore, it may be safe to say that the $SU(N)$ SYM$_4$ also provides the other example for which we can define a reliable stochastic process that has a unique equilibrium limit. However, our interest is to take the continuum limit by taking $N \rightarrow \infty$. In the large N limit, we expect that there exists a spontaneous breakdown of the $U(1)^d$ symmetry, and the space-time may collapse in the reduced model of $U(N)$ YM theory. The reduced model of $U(N)$ SYM$_4$ may become a large N limit of the continuum SYM$_4$: there is no non-trivial double scaling limit to a superstring theory. Unfortunately, it seems that we have to leave the verification of the existence of a unique equilibrium limit for the stochastic process defined by the physically interesting cases for a later work.  

As a regularization procedure in the superfield formalism for SYM$_4$, we have assumed the following prescription for the superspace $\delta$-function: 
\bea
\delta^8 (0) 
\equiv 
\delta^4 (x-x') \delta^2 (\theta - \theta') \delta^2 ({\bar \theta} - {\bar \theta}') \Big|_{x=x', \theta = \theta',\ {\bar \theta} = {\bar \theta}'} = 0 .
\ena
This prescription is indicated by the regularization through dimensional reduction. In general, in the SQM approach via It${\bar {\rm o}}$ calculus, the origin of anomalies is traced to the contact term,\cite{NOTY} and we have to regularize $\delta^8 (0)$ with a suitable regularization procedure. In SYM$_4$, there exists a superconformal anomaly which, in the regularization through dimensional reduction, comes from the counterterm in the superfield formalism. Therefore, the prescription is consistent with the origin of the anomaly in this case.    
For SYM$_{10}$, by contrast, we can keep the contact term that is proportional to the space-time $\delta^{10} (0)$. It is possible to derive the Ward-Takahashi identity for the Yang-Mills gauge current, and there exists the gauge anomaly, as expected from the contact term, if we adopt the heat kernel regularization to evaluate the contact term of the fermionic noise variables in the background gauge field. It is not clear, however, whether the regularization procedure is consistent with the global supersymmetry. The regularization with the heat kernel method may also be useful to regularize the superspace $\delta^8 (0)$ in the superspace formalism for SYM$_4$. In this case, we can keep the contact term with the heat kernel regularization. This regularization procedure may be preferable for the purpose of further investigation of the correspondence of SYM$_4$ in the superfield formalism to LGT$_d$. 

As we have briefly illustrated in \S 3, 
it is necessary to introduce the stochastic gauge fixing procedure for the perturbative analysis with gauge variant quantities. This is formulated by introducing a chiral and an anti-chiral superfield, as the Zwanziger gauge fixing functions, and specifying them as functions of the vector superfield $V$. 
It is then possible to demonstrate the non-renormalization theorem and the perturbative renormalizability for SYM$_4$ in the superfield formalism in terms of the BRST invariant stochastic action principle. It has also been shown that the SQM approach is formally equivalent to the Faddeev-Popov prescription in the path-integral approach by specifying the Zwanziger gauge fixing functions\cite{Nakazawa3}. In order to support the formal proof, we have also carried out a one-loop perturbation by applying  the background field method. The one-loop $\beta$-function for the gauge coupling is identical to that given in the path-integral method.\cite{Nakazawa4} This supports the equivalence of the stochastic gauge fixing procedrue and the Faddeev-Popov prescription.

One of our main interests is to apply SQM to large N reduced models and to construct collective field theories of Wilson loops, as illustrated in the naive zero volume limit of the YM case\cite{EKN}. By developing the SQM approach, we hope to obtain some hints for the construction of a manifestly Lorentz invariant formulation of string field theories, in praticular, for IIB superstring field theory. It seems also to be possible to evaluate the expectation values of Wilson loops in terms of the Langevin equations. To do this, we have to clarify the structure of the ground states in colletive field theories of large N reduced models. Regarding the application of SQM to the IIB matrix model, since the Pfaffian is complex, we may encounter some difficulties in defining a reliable stochastic process using the Langevin equation. Another way to give a physical meaning in the SQM approach is to interpret the collective field theory Hamiltonian as a Hamiltonian constraint for the IIB superstring field theory. This interpretation was also given in the prototype. It is an open question whether the collective field theory Hamiltonian constraint can be reduced to string field theory Hamiltonians in non-covariant gauges such as the light-cone gauge.

\section*{Acknowledgements}

The author would like to thank S. Iso, S. Komine, N. Okada and T. Suyama for enlightening discussions and all the members of the theory group at KEK for their hospitality. 
This work was supported in its early stages by the Ministry of Education, Calture, Sports, Science and Technology of Japan, through a Grant-in-Aid for Scientific Research (B) ( No.13135216 ).

\appendix

\section{Derivation of the Fokker-Planck Equation}\label{sec:A}

In this appendix, we derive the Fokker-Planck equation. To do this, we define a supersymmetric local gauge invariant path-integral measure and introduce a probability density in the superfield formalism. 

Let us introduce a differential operator ${\hat E}_a (z)$ defined by 
\bea
\label{eq:left-derivative-eq1} 
\displaystyle{\frac{\pa}{\pa V^a (z)}} e^{2V (z)} 
& = & 
\displaystyle{\frac{2}{{\rm k}}}\int^1_0 ds {\rm Tr}\Big( 
e^{2sV (z)}t_a e^{-2sV (z)} t^b 
\Big) t_b e^{2V (z)} 
\equiv 
2 K_a^{\ b} (z) t_b e^{2V (z)}             \ , \nn 
{\hat E}_a (z) 
& \equiv & 
\displaystyle{\frac{1}{2}} L_a^{\ b} (z) \displaystyle{\frac{\pa}{\pa V^b (z)}}      \ , \quad 
K_a^{\ c} (z) L_c^{\ b} (z) 
 =  
L_a^{\ c} (z) K_c^{\ b} (z) = \delta_a^{\ b}      \ . 
\ena 
Here, $z$ denotes the set of coordinates $( x, \theta, {\bar \theta} )$ in superspace. The coefficients $L_a^{\ b} (z)$ and $K_a^{\ b} (z)$ satisfy $L_a^{\ b} (z)^\dagger = L^b_{\ a} (z) $ 
and 
$K_a^{\ b} (z)^\dagger = K^b_{\ a} (z) $. 

By definition, we have 
\bea
\label{eq:left-derivative-eq2}
{\hat E}_a (z) e^{2V (z)} 
& = & 
t_a e^{2V (z)}        \ , \nn 
\Big[ {\hat E}_a (z), {\hat E}_b (z) \Big] 
& = & 
-if_{ab}^{\ \ c} {\hat E}_c (z)      \ . 
\ena
The coefficients $L_a^{\ b} (z)$ and $K_a^{\ b} (z)$ also satisfy Maurer-Cartan type equations: 
\bea 
\label{eq:Maurer-Cartan}
L_a^{\ c} (z) \pa_c L_b^{\ d} (z) - L_b^{\ c} (z) \pa_c L_a^{\ d} (z) 
& = & 
- 2 i f_{ab}^{\ \ c}L_c^{\ d} (z)       \ , \nn  
\pa_b K_c^{\ a} (z) - \pa_c K_b^{\ a} (z) 
& = & 
+ 2 i f_{b'c'}^{\ \ a}K_b^{\ b'} (z) K_c^{\ c'} (z)     \ . 
\ena 
We can also introduce an alternative definition of the differential operator ${\hat E}^{(R)}_a (z)$: 
\bea
\label{eq:right-derivative-eq1} 
{\hat E}^{(R)}_a (z) 
& \equiv & 
L^b_{\ a} (z) \displaystyle{\frac{\pa}{\pa V^b (z)}}      \ , \nn
{\hat E}^{(R)}_a (z) e^{2V (z)} 
& = & 
e^{2V (z)} t_a                  \ , \nn 
\Big[ {\hat E}^{(R)}_a (z), {\hat E}^{(R)}_b (z) \Big] 
& = & 
+ if_{ab}^{\ \ c} {\hat E}^{(R)}_c (z)      \ . 
\ena 

Although $U(z) \equiv e^{2V (z)}$ is not a group element, the differential operators ${\hat E}_a (z)$ and ${\hat E}^{(R)}_a (z)$ are analogues of the left and right Lie derivatives on the group manifold, respectively. As we need the functional derivative, we define the following: 
\bea
\label{eq:left-derivative-eq3}
{\hat {\cal E}}_a  
\equiv 
\displaystyle{\frac{1}{2}}
\int dz L_a^{\ b} (z) \displaystyle{\frac{\delta\ \ }{\delta V^b (z)}} 
\equiv 
\int dz {\hat {\cal E}}_a (z)   \ . 
\ena
The functional derivative with respect to the vector superfield is defined by 
\bea
\displaystyle{\frac{\delta V^b (z') }{\delta V^a (z)}} 
& = & 
\delta ( z - z' ) \delta_a^{\ b}       \ , \nn
\delta ( z - z' ) 
& \equiv &
\delta^2 ( \theta - \theta' ) 
\delta^2 ( {\bar \theta} - {\bar \theta}' ) \delta^4 ( x - x' )     \ .
\ena
The differential operator ${\hat {\cal E}}_a$ also satisfies the relations 
\bea
\label{eq:left-derivative-eq4}
{\hat {\cal E}}_a e^{2V (z)}  
& = & 
t_a e^{2V (z)}        \  ,    \nn   
\Big[ {\hat {\cal E}}_a, {\hat {\cal E}}_b \Big]  
 & = & 
-if_{ab}^{\ \ c} {\hat {\cal E}}_c      \  . 
\ena

Let $F[ U ]$ be a functional of $U (z)$. Then the time evolution of $F[ U ]$ is given by,
\bea
\label{eq:Fokker-Planck-appendix-eq1}
\Delta F[ U ] 
& = & 
\int\!\!\! dz \Delta U ( z )_{ij} \displaystyle{\frac{\delta F}{\delta U (z)_{ij}}} 
+ \displaystyle{\frac{1}{2}}\int\!\!\! dz dz' \Delta U ( z )_{ij} \Delta U ( z' )_{kl} \displaystyle{\frac{\delta^2 F}{\delta U (z)_{ij} \delta U (z')_{kl}}}   \ , \nn
& = & 
\int\!\!\! dz \Big( 
- \Delta \t X (z) + \Delta w (z)
\Big) U(z) \displaystyle{\frac{\delta F}{\delta U (z)}}    \nn 
& {} & \qquad
+ \displaystyle{\frac{1}{2}}\int\!\!\! dz dz' \Big( \Delta w (z) U ( z ) \Big) \Big( \Delta w ( z' ) U ( z' ) \Big) \displaystyle{\frac{\delta^2 F}{\delta U (z) \delta U (z')}}   \ , \nn
& = & 
- \Delta \t \int\!\!\! dz \Big( 
X (z) U(z)
\Big)_{ij}  \displaystyle{\frac{\delta F}{\delta U (z)_{ij}}}    \nn 
& {} & \qquad
+ \Delta \t \int\!\!\! dz \Big( {\hat E}^a(z) U ( z ) \Big)_{ij} \Big( {\hat E}_a ( z ) U ( z ) \Big)_{kl} \displaystyle{\frac{\delta^2 F}{\delta U (z)_{ij} \delta U (z)_{kl}}}   \ , \nn
& = & 
 \Delta \t \int\!\!\! dz \Big(  
- X (z)^a {\hat {\cal E}}_a(z) 
+  {\hat {\cal E}}^a (z) {\hat {\cal E}}_a ( z )   
\Big) F[ U ]         \ .
\ena 

In order to derive the Fokker-Planck equation, we define the probability distribution $P( \t, U )$ as 
\bea
\label{eq:Fokker-Planck-appendix-eq2}
\langle  F[ U( \t ) ] \rangle_{\Delta w}
= \int\!\!\!  F[ U ]P( \t, U ) \sqrt{G}{\cal D}V \ . 
\ena
In this expression, the measure of the integral is defined as follows. Let us introduce a metric defined by 
$G^{ab} (z) \equiv {1\over 4}L_c^{\ a} (z) L_c^{\ b} (z) = {1\over 4}L^a_{\ c} (z) L^b_{\ c} (z)$, 
$G_{ab} (z) \equiv 4 K_a^{\ c} (z) K_b^{\ c} (z) = 4 K^c_{\ a} (z) K^c_{\ b} (z)$ 
and 
$G (z) \equiv {\rm det}G_{ab} (z)$. We note that 
$G_{ab}^\dagger = G_{ab}$; that is, the metric is real. 
The integration measure is defined by 
\bea
\label{eq:integration-measure}
d\mu ( U ) 
& \equiv & 
\sqrt{G}{\cal D}V        \ , \nn
\sqrt{G} 
& = & 
\prod_{x} \sqrt{G(z)}    \ , \nn
{\cal D}V 
& = & 
\prod_{x} dC(x)d\chi(x)d{\bar \chi}(x)dv(x)d\lambda(x)d{\bar \lambda}(x)dD(x)  \ .
\ena
This definition is similar to that of the Haar measure on a group manifold. We give a proof of the invariance of the integration measure under the stochastic-time $independent$ local gauge transformation 
\bea
\label{eq:local-gauge-tr-appendix-eq1}
U
& \rightarrow & 
e^{-i\Lambda^\dagger} U e^{i\Lambda}     \ , \nn 
U^{-1}
& \rightarrow & 
e^{-i\Lambda} U^{-1} e^{i\Lambda^\dagger}     \ , 
\ena
where $\Lambda$ and $\Lambda^\dagger$ are $SU(N)$ algebra-valued chiral superfields: 
$
{\bar D}_{\dot \alpha} \Lambda = D_\alpha \Lambda^\dagger = 0 \ . 
$
The local gauge transformation is expressed in terms of the vector superfield $V (z)$ as 
\bea
\label{eq:local-gauge-tr-appendix-eq2}
\delta V^a (z) 
= 
- \displaystyle{\frac{i}{2}} ( \Lambda^\dagger (z))^b L_b^{\ a} (z)
+ \displaystyle{\frac{i}{2}} L^a_{\ b} (z) \Lambda^b (z)          \ .
\ena
From (\ref{eq:Maurer-Cartan}), the variation on the vector superfield satisfies the Killig vector equation: 
\bea
\label{eq:Killing-vector-eq1}
\delta G_{ab} 
& \equiv & 
-i ( \Lambda^\dagger )^c {\hat E}_c G_{ab} 
+ i \Lambda^c {\hat E}^{(R)}_c G_{ab}    \ , \nn
& = & 
\delta V^c \displaystyle{\frac{\pa}{\pa V^c}} G_{ab} \ , \nn 
& = & 
\delta G_{ac} \displaystyle{\frac{\pa ( \delta V^b )}{\pa V^c}} 
+ \delta G_{cb} \displaystyle{\frac{\pa ( \delta V^a )}{\pa V^c}}  \ .
\ena
This implies that the integration measre is transformed by a total derivative under the local gauge transformation. By changing the integration variable through the local gauge transformation, we obtain 
\bea
\label{eq:Fokker-Planck-appendix-eq3}
\delta \int\!\!\!  F[ U ]P( \t, U ) \sqrt{G}{\cal D}V  
& = & 
- i \int\!\!\! {\cal D}V  \int\!\!\! dz  {\hat {\cal E}}_c (z)
\Big( ( \Lambda^\dagger (z) )^c F[ U ]P( \t, U ) \sqrt{G}   \Big)   \nn
& {} & \quad 
+ i \int\!\!\! {\cal D}V \int\!\!\! dz  {\hat {\cal E}}^{(R)}_c (z)
\Big( \Lambda^c (z) F[ U ]P( \t, U ) \sqrt{G}   \Big)   \ , \nn
& = & 0    \ .
\ena
Hence, we conclude that the definition of the probabilitiy distribution is local gauge invariant. It is also confirmed that we can integrate by parts with respect to the derivative operations ${\hat {\cal E}}_a (z)$ and $ {\hat {\cal E}}^{(R)}_a (z)$, defined in (\ref{eq:left-derivative-eq3}). 

Now, it is easy to derive the Fokker-Planck equation. It is given by 
\bea
\label{eq:Fokker-Planck-appendix-eq4}
\displaystyle{\frac{\pa\ }{\pa \t}}P( \t, U ) 
= 
\int\!\!\! dz {\hat {\cal E}}_a(z) \Big(  
{\hat {\cal E}}^a (z) + X^a (z)  
\Big) P( \t, U )         \ .
\ena 
To reproduce  ${\cal N}$=1 SYM$_4$ in the equilibrium limit, we simply choose 
\bea
\label{eq:Fokker-Planck-appendix-eq5}
X^a (z) = {\hat {\cal E}}^a (z) S   \ , 
\ena
with $S$ in (\ref{eq:4Daction}). 
It is also clear that we need the path-integral measure (\ref{eq:integration-measure}) even in the equilibrium limit for the definition of the local gauge invariant partition function in the path-integral formalism: 
\bea
\label{eq:super-gauge-inv.-partition-fun.}
Z \equiv \int\!\!\! e^{- S} \sqrt{G} {\cal D}V  \ .
\ena

To end this appendix, we check the invariance of the Fokker-Planck equation under the local gauge transformation. 
Since 
$
{\hat {\cal E}}(z) S 
$ 
is transformed as 
$
 {\hat {\cal E}}(z) S  
\rightarrow e^{-i\Lambda\dagger} \Big(
 {\hat {\cal E}}(z) S  \Big)
  e^{i\Lambda\dagger} ,\ 
$ 
the differential operator ${\hat E}$ must be transformed as 
$
{\hat E}
\rightarrow e^{-i\Lambda\dagger} {\hat E} e^{i\Lambda\dagger} 
$. 
Under this transformation, we have 
$
{\hat E}'_a e^{2V'} = {\tilde t}_a e^{2V'}  \ , 
$
where by definition 
$
{\hat E}'_a  =  {\hat E}_a  \ 
$
and 
$
{\tilde t}_a  = e^{-i\Lambda\dagger} t_a e^{i\Lambda\dagger} \ .
$
The transformed elements of the $SU(N)$ algebra also satisfy 
$
[ {\tilde t}_a, {\tilde t}_b] = if_{ab}^{\ \ c} {\tilde t}_c ,\ ({\tilde t}^a)_{ij} ({\tilde t}_a)_{kl} = {\rm k} ( \delta_{il}\delta_{kj} 
- \displaystyle{\frac{1}{N}}\delta_{ij}\delta_{kl} ) .    \
$
Hence the Fokker-Planck equation is invariant. 


\section{A Kerneled Form of the Superfield Langevin Equation \\ 
and the Almost Wess-Zumino Gauge for the Component Expansion}\label{sec:B}

In this appendix, we clarify that the superfield Langevin equation includes a kernel written in terms of the superfield. In terms of the kerneled form of the superfield Langevin equation, we point out that the Wess-Zumino gauge is impossible in the superfield Langevin equation. In order to decompose the superfield Langevin equation into its components, we redefine the vector superfield so that the auxiliary component fields appear in the superfield Langevin equation only through their derivatives with respect to the stochastic time. We refer to (\ref{eq:4DLangevin-B-eq5}) as the superfield Langevin equation in the \lq\lq almost Wess-Zumino gauge \rq\rq. 

Let us introduce the Lie derivative $L_V$, which is defined by $L_V X = [V, X]$. If we define the inner product $(X\cdot Y) = \displaystyle{\frac{1}{{\rm k}}}{\rm Tr} (XY)$, then we have $(X\cdot L_V Y) = (Y\cdot (L_V)^t X) = - (Y\cdot L_V X)$. This also satisfies the relation 
$
[\ L_V,\ L_{V'} ] 
= L_{[V,\ V' ]}
$. 
By introducing the notation $(L_V)_a^{\ b} \equiv \Big( t_a\cdot L_V t^b \Big)$, we define the operators 
\bea
\label{eq:Maurer-Cartan-coefficients}
K 
& \equiv & 
 \displaystyle{\frac{1}{2 L_V}}( 1 - e^{-2L_V} )     \ , \nn
L 
& \equiv & 
 \displaystyle{\frac{2 L_V}{ 1 - e^{-2L_V}}}    \ ,
\ena
which correspond to the coefficients $K_a^{\ b}$ and $L_a^{\ b}$, respectively. In this notation, the two equivalent Langevin equations (\ref{eq:4DLangevin-eq4}) and (\ref{eq:4DLangevin-eq6}) are expressed as follows. 
(\ref{eq:4DLangevin-eq4}) becomes 
\bea
\label{eq:4DLangevin-B-eq1}
\displaystyle{\frac{(e^{2L_V} - 1 )}{ L_V}}\Delta V ( \t, z ) 
=  \Delta \t \displaystyle{\frac{1}{4g^2}}  \Big( 
e^{2L_V}  {\cal D}{\cal W} + {\bar {\cal D}}{\bar {\cal W}} 
\Big) + \Delta w ( \t, z )          \ ,  
\ena
and (\ref{eq:4DLangevin-eq6}) becomes 
\bea
\label{eq:4DLangevin-B-eq2}
\Delta V ( \t, z )
& = & 
 \Delta \t \displaystyle{\frac{1}{4g^2}}  \Big( 
( 1 - e^{-2L_V} )^{-1} L_V {\cal D}{\cal W}      \nn
& {} & \qquad \qquad \qquad 
+ (e^{2L_V} - 1 )^{-1} L_V {\bar {\cal D}}{\bar {\cal W}} 
\Big) + \Delta_w \Xi ( \t, z ) \ ,  \nn 
\Delta_w \Xi ( \t, z ) 
& \equiv & 
( e^{2L_V} - 1 )^{-1} L_V \Delta w  ( \t, z )    \ .
\ena
Here, we have defined 
\bea
\label{eq:field-strength}
{\cal D}{\cal W} 
& \equiv & 
D^\alpha W_\alpha + \{ 
W_\alpha , e^{-2V} D^\alpha e^{2V} 
\}                \ , \nn
{\bar {\cal D}}{\bar {\cal W}} 
& \equiv & 
{\bar D}_{\dot \alpha} {\bar W}^{\dot \alpha} + \{ 
{\bar W}^{\dot \alpha} , e^{2V} {\bar D}_{\dot \alpha} e^{-2V} 
\}              \ .
\ena
From these expression, the equivalence is trivial. We also confirm the hermiticity of the Langevin equations. The hermitian conjugation of (\ref{eq:4DLangevin-B-eq1}) ((\ref{eq:4DLangevin-eq4})) is given by 
\bea
\label{eq:4DLangevin-B-eq3}
\displaystyle{\frac{(1 - e^{-2L_V}  )}{L_V}} \Delta V^\dagger ( \t, z ) 
=  \Delta \t \displaystyle{\frac{1}{4g^2}}  \Big( 
e^{- 2L_V} {\bar {\cal D}}{\bar {\cal W}} + {\cal D}{\cal W} 
\Big) + \Delta w^\dagger ( \t, z )          \ .  
\ena
The hermiticity assignment of the noise superfield gives 
\bea
\label{eq:noise-hermiticity}
\Delta w^\dagger ( \t, z )  = e^{- 2L_V} \Delta w ( \t, z )   \ . 
\ena
By multiplying (\ref{eq:4DLangevin-B-eq3}) by $e^{2 L_V}$, we find that $\Delta V ( \t, z )^\dagger$ satisfies the same Langevin equation as $\Delta V ( \t, z )$. Hence, 
$\Delta V( \t, z )^\dagger = \Delta V ( \t, z ) $ 
holds 
if the time evolution begins with the initial condition $V^\dagger ( 0, z ) = V ( 0, z )$. 

(\ref{eq:4DLangevin-B-eq2}) ((\ref{eq:4DLangevin-eq6})) can also be expressed in the form 
\bea
\label{eq:4DLangevin-B-eq4}
\Delta V ( \t, z )
= 
- \Delta \t 
 ( e^{2L_V} - 1 )^{-1} L_V ( 1 - e^{-2L_V} 
 )^{-1} L_V \displaystyle{\frac{\delta S}{\delta V^t}}   
+ \Delta_w \Xi ( \t, z ) \ . 
\ena
This means that we choose the superfield kernel 
$ (e^{2L_V} - 1 )^{-1} L_V ( 1 - e^{-2L_V} )^{-1} L_V  $ 
for the Langevin equation in the superfield formalism. We note that this gives simply the factor ${1\over 4}$ in the U(1) limit. 

We next discuss the Wess-Zumino gauge for the superfield Langevin equation.  In order to impose the Wess-Zumino gauge condition, 
$
C = \chi = {\bar \chi} = M = N = 0 
$, 
the chiral superfields $\Lambda$ and $\Lambda^\dagger$ are given as follows in the linear approximation of the local gauge transformation 
$
e^{2V'} 
= 
e^{- i \Lambda^\dagger}e^{2V} e^{ i \Lambda}     
$: 
\bea
\label{eq:Wess-Zumino-B-eq1}
\Lambda (z) 
& = & 
\Lambda ( y, \theta )    \ , \nn 
& = & 
 i C(y) - 2 \theta \chi (y) - \theta^2 ( M(y) + i N (y))  \ , \nn
\Lambda^\dagger (z) 
& = & 
\Lambda^\dagger ( y^\dagger, {\bar \theta} )    \ , \nn
& = & 
 - i C( y^\dagger ) - 2 {\bar \theta} {\bar \chi} ( y^\dagger ) - {\bar \theta}^2 ( M( y^\dagger ) - i N ( y^\dagger ))   \ ,
\ena
where 
$y^m = x^m + i\theta\sigma^m {\bar \theta}$ and 
${y^m}^\dagger = x^m - i\theta\sigma^m {\bar \theta}$. 
This is not sufficient, because the transformation is highly nonlinear. Suppose that the chiral superfields $\Lambda$ and $\Lambda^\dagger$ are determined completely to gauge away the auxiliary component fields. Then we obtain a gauge transformed form of the vector superfield from that in the Wess-Zumino gauge. 
Explicitly, we have 
\bea
\label{eq:Wess-Zumino-B-eq2}
e^{2V} 
& \equiv & 
e^{+ i \Lambda^\dagger}e^{2V |_{{\rm WZ}}} e^{ - i \Lambda}\ , \nn 
V |_{{\rm WZ}} (z)
& = & 
- \theta\sigma^m {\bar \theta} v_m (x)        
 +
i \theta^2 {\bar \theta}  {\bar \lambda} (x)         
- i {\bar \theta}^2 \theta \lambda (x) 
+ \displaystyle{\frac{1}{2}}\theta^2 {\bar \theta}^2  D (x) \ . 
\ena

The residual component of the chiral superfields for the local gauge transformation is now given by $ a (y) $, where $a (x)$ is a real function, but 
$a (y)^\dagger = a (y^\dagger )$. 
The transformation property of the vector superfield (\ref{eq:Wess-Zumino-B-eq2}) under this residual gauge transformation is given by 
\bea
\label{eq:Wess-Zumino-B-eq3}
e^{2V'} 
& \equiv & 
e^{- i a( y^\dagger )}e^{2V}  e^{ i a(y) } \ , \nn 
& = &
e^{- i a( y^\dagger )}e^{+ i \Lambda^\dagger} e^{i a (y^\dagger )}e^{- i a( y^\dagger )}e^{2V |_{{\rm WZ}}} e^{ i a (y)}e^{- i a (y)}e^{ - i \Lambda } e^{ i a(y) }    \ , \nn
& \equiv & 
e^{+ i {\Lambda'}^\dagger}e^{2V |_{{\rm WZ}}'} e^{ - i \Lambda'}             \ . 
\ena
This defines the residual local gauge symmetry for the components in the Wess-Zumino gauge, i.e., 
$
v_m (x) \rightarrow v_m (x) + {\cal D}_m a (x)   \ , 
\lambda (x) \rightarrow  \lambda (x) + i[ \lambda (x),\ a(x) ] 
$ 
and 
$
D (x) \rightarrow  D (x) + i[ D (x),\ a(x) ] .
$
It also rotates the auxiliary component fields 
$\Lambda'( y, \theta ) 
= e^{- i a (y)} \Lambda ( y, \theta )e^{ i a(y) } $. 

In order to obtain an expression for the superfield Langevin equation that is convenient to decompose into component fields, we substitute the definition of superfield (\ref{eq:Wess-Zumino-B-eq2}) into (\ref{eq:4DLangevin-B-eq1}). We have shown that the Langevin equation is covariant under the stochastic-time $independent$ local gauge transformation, while (\ref{eq:Wess-Zumino-B-eq2}) is the stochastic-time $dependent$ local gauge transformation due to the stochastic-time dependence of the chiral superfield $\Lambda ( \t, z )$, which represents the auxiliary degrees of freedom in the vector multiplet. The r.h.s. of (\ref{eq:4DLangevin-B-eq1}) is covariant; that is  
\bea
\label{eq:Wess-Zumino-B-eq4}
 \Big( 
e^{2L_V}  {\cal D}{\cal W} + {\bar {\cal D}}{\bar {\cal W}} 
\Big)            
& = & 
e^{+ i \Lambda^\dagger} 
\Big( 
e^{{2L_V} |_{{\rm WZ}}}  {\cal D}{\cal W}|_{{\rm WZ}} + {\bar {\cal D}}{\bar {\cal W}}|_{{\rm WZ}}    
\Big) e^{ - i \Lambda^\dagger}     \ ,  \nn 
\Delta w ( \t, z )         
& \equiv &  
e^{+ i \Lambda^\dagger} \Delta w |_{{\rm WZ}} ( \t, z )  e^{ - i \Lambda^\dagger}        \ .
\ena
We note that 
$e^{2L_V} = e^{+ i L_{\Lambda^\dagger}} e^{{2L_V} |_{{\rm WZ}}}e^{- i L_\Lambda} $. 
As it is clear from the definition,  $\Delta w |_{{\rm WZ}} ( \t, z )$ satisfies the same correlation relation as $\Delta w ( \t, z )$. 
As a consequence, the correlation between $\Delta w |_{{\rm WZ}}$ and $\Delta w |_{{\rm WZ}}^\dagger$ is given by 
\bea
\label{eq:Wess-Zumino-nise-B-eq1}
\langle 
\Delta w |_{{\rm WZ}} ( \t, z ) \Delta w |_{{\rm WZ}}^\dagger ( \t, z' ) 
\rangle_{\Delta w_\t} 
= 2{\rm k} \Delta \t \Big( \delta_{il}\delta_{kj} - \displaystyle{\frac{1}{N}} \delta_{ij}\delta_{kl} \Big) \delta ( z-z' )        \ .
\ena
However, this does not mean that the noise superfield is a vector superfield. 
The hermiticity condition is also given by  
\bea
\label{eq:Wess-Zumino-nise-B-eq2}
\Delta w |_{{\rm WZ}}^\dagger ( \t, z ) = e^{-{2L_V} |_{{\rm WZ}}} \Delta w |_{{\rm WZ}} ( \t, z )    \ .
\ena

The l.h.s. of the Langevin equation (\ref{eq:4DLangevin-B-eq1}) is $not$ $covariant$ under the stochastic-time dependent local gauge transformation. The l.h.s. reads 
\bea
\label{eq:Wess-Zumino-B-eq5}
( \Delta e^{2V} )e^{- 2V} 
& = & 
\displaystyle{\frac{(e^{2L_V} - 1 )}{ L_V}}\Delta V    \ , \nn 
& = & 
\displaystyle{\frac{(e^{iL_{\Lambda^\dagger}} - 1 )}{ L_{\Lambda^\dagger}}}\Delta \Lambda^\dagger 
+ e^{iL_{\Lambda^\dagger}} \displaystyle{\frac{(e^{2L_V|_{\rm WZ}} - 1 )}{ L_V|_{\rm WZ}}}\Delta V|_{\rm WZ}   \nn 
& {} & \qquad 
+ e^{iL_{\Lambda^\dagger}} e^{2L_V|_{\rm WZ}} \displaystyle{\frac{(e^{-i L_\Lambda} - 1 )}{ L_\Lambda}}\Delta \Lambda    \ . 
\ena
Since the Langevin equation is not covariant under the stochastic-time dependent local gauge transformation, the auxiliary component fields cannot be gauged away. This means that it is impossible to take the Wess-Zumino gauge in the superfield Langevin equation. In order to simplify its expression, we redefine the time development of the auxiliary superfields that does not include the components in the Wess-Zumino gauge, 
\bea
\label{eq:Wess-Zumino-B-eq6}
\Delta \Theta^\dagger 
& \equiv & 
\displaystyle{\frac{( 1 - e^{- iL_{\Lambda^\dagger}} )}{ L_{\Lambda^\dagger}}}\Delta \Lambda^\dagger        \ , \nn 
\Delta \Theta
& \equiv & 
\displaystyle{\frac{(e^{-i L_\Lambda} - 1 )}{ L_\Lambda}}\Delta \Lambda        \ . 
\ena
These collective superfields are also chiral:  
\bea
\label{eq:Wess-Zumino-B-eq7}
{\bar D}_{\dot \alpha}  \Delta \Theta
& = & 
D_\alpha \Delta \Theta^\dagger  = 0 \ , \nn
\Delta \Theta ( y, \theta)^\dagger 
& = & 
\Delta \Theta^\dagger ( y^\dagger, {\bar \theta} )     \  . 
\ena
They may be expanded as 
\bea
\label{eq:Wess-Zumino-B-eq8}
\Delta \Theta (z) 
& = & 
\Delta \Theta ( y, \theta )                              \ , \nn 
& \equiv & 
  \Delta {\tilde C} (y) + 2i \theta \Delta {\tilde \chi} (y) + i \theta^2 ( \Delta {\tilde M} (y) + i \Delta {\tilde N} (y))                                    \ , \nn
\Delta \Theta^\dagger (z) 
& = & 
\Delta \Theta^\dagger ( y^\dagger, {\bar \theta} )       \ , \nn
& \equiv & 
 \Delta {\tilde C} ( y^\dagger ) - 2i {\bar \theta} \Delta {\bar {\tilde \chi}} ( y^\dagger ) - i {\bar \theta}^2 ( \Delta {\tilde M} ( y^\dagger ) - i \Delta {\tilde N} ( y^\dagger ))                            \ . 
\ena
Therefore, by using the definition, auxiliary component fields appear in the superfield Langevin equation only through their derivatives with respect to the stochastic time. 
The superfield Langevin equation in the \lq\lq almost Wess-Zumino gauge\rq\rq is given by 
\bea
\label{eq:4DLangevin-B-eq5}
\Delta V|_{\rm WZ}   
& + & 
\displaystyle{\frac{L_V|_{\rm WZ}}{(e^{2L_V|_{\rm WZ}} - 1 )}} 
\Delta \Theta^\dagger 
+ \displaystyle{\frac{L_V|_{\rm WZ}}{( 1 - e^{- 2L_V|_{\rm WZ}} )}} 
 \Delta \Theta                     \nn 
& = & 
 \Delta \t \displaystyle{\frac{1}{4g^2}}  \Big\{ 
\displaystyle{\frac{L_V|_{\rm WZ}}{( 1 - e^{- 2L_V|_{\rm WZ}} )}}  {\cal D}{\cal W}|_{\rm WZ}                                \nn 
& {} & \qquad 
+ \displaystyle{\frac{L_V|_{\rm WZ}}{(e^{2L_V|_{\rm WZ}} - 1 )}}  {\bar {\cal D}}{\bar {\cal W}}|_{\rm WZ} 
\Big\} + \Delta_w \Xi|_{\rm WZ}                                                             \ ,  \nn 
\Delta_w \Xi|_{\rm WZ} 
& \equiv & 
\displaystyle{\frac{L_V|_{\rm WZ}}{(e^{2L_V|_{\rm WZ}} - 1 )}}  \Delta w|_{\rm WZ}     \ .
\ena
We note that, though all the auxiliary component fields remain, the time evolution of these auxiliary component fields, 
(${\tilde C}$, ${\tilde \chi}$, ${\tilde {\bar \chi}}$, ${\tilde M}$, ${\tilde N}$), is completely determined by the components in the Wess-Zumino gauge, i.e., ($v_m$, $\lambda$, ${\bar \lambda}$, $D$).


\section{Derivation of Langevin Equations for Component Fields}\label{sec:C}

In this appendix, we derive the Langevin equations for the component fields of SYM$_4$ in the superfield formalism. The component expansion is obtained for (\ref{eq:4DLangevin-B-eq5}). 

For the time development of the auxiliary component fields, (${\tilde C}$, ${\tilde \chi}$, ${\tilde {\bar \chi}}$, ${\tilde M}$, ${\tilde N}$), (\ref{eq:4DLangevin-B-eq5}) can be reduced to 
\bea
\label{eq:4DLangevin-c-eq1}
\Delta \Theta^\dagger 
& = & 
\Delta \t \displaystyle{\frac{1}{4g^2}} 
{\bar {\cal D}}{\bar {\cal W}}|_{\rm WZ} 
+ \displaystyle{\frac{1}{2}} \Delta w|_{\rm WZ}             \ ,  \nn 
 \Delta \Theta                     
& = & 
 \Delta \t \displaystyle{\frac{1}{4g^2}} 
 {\cal D}{\cal W}|_{\rm WZ}                                
+ \displaystyle{\frac{1}{2}} \Delta w^\dagger|_{\rm WZ}                 \ .
\ena
Here, we have used the hermiticity condition (\ref{eq:Wess-Zumino-nise-B-eq2}). 
The reduction is not correct for the components proportional to $\theta\sigma^m{\bar \theta}$, $\theta {\bar \theta}^2$, ${\bar \theta}\theta^2$ and $\theta^2 {\bar \theta}^2$. For (${\tilde C}$, ${\tilde \chi}$, ${\tilde {\bar \chi}}$, ${\tilde M}$, ${\tilde N}$), (\ref{eq:4DLangevin-c-eq1}) reads 
\bea
\label{eq:4DLangevin-c-eq2}
\Delta {\tilde C} 
& = & 
- \displaystyle{\frac{\Delta \t}{2}} \displaystyle{\frac{1}{g^2}} D  
+ \displaystyle{\frac{1}{2}} \Delta w^\dagger|       \ , \nn
\Delta {\tilde \chi} 
& = & 
- i \displaystyle{\frac{\Delta \t}{2}} \displaystyle{\frac{1}{g^2}} \sigma^m {\cal D}_m {\bar \lambda}  
-  \displaystyle{\frac{i}{2}} \Delta w^\dagger|_{\theta}       \ , \nn
\Delta {\bar {\tilde \chi}} 
& = &   
- i \displaystyle{\frac{\Delta \t}{2}} \displaystyle{\frac{1}{g^2}} {\bar \sigma}^m {\cal D}_m \lambda 
+ \displaystyle{\frac{i}{2}} \Delta w |_{\bar \theta}       \ , \nn    
 \Delta {\tilde M} + i \Delta {\tilde N} 
& = &  
 -  i \Delta w^\dagger|_{\theta^2}       \ , \nn
 \Delta {\tilde M} - i \Delta {\tilde N} 
& = & 
 i \Delta w |_{{\bar \theta}^2}       \ . 
\ena
The components of the noise superfields $W (z) \equiv ( \Delta w (z),\ \Delta w^\dagger (z) )$ are given by the following definition (where we have suppressed the suffix \lq\lq $|_{\rm WZ}$ \rq\rq for the noise superfield for simplicity): 
\bea
W 
& = & 
 W | + \theta  W |_{\theta} + {\bar \theta} W |_{{\bar \theta}} + \theta^2 W |_{\theta^2} + {\bar \theta}^2\ W |_{{\bar \theta}^2}      \nn
& {} & \quad 
+ \theta\sigma^m{\bar \theta} W |_{\theta\sigma^m{\bar \theta}} + {\bar \theta}^2 \theta W |_{{\bar \theta}^2 \theta } + \theta^2 {\bar \theta} W |_{\theta^2 {\bar \theta}} + \theta^2{\bar \theta}^2\ W |_{\theta^2{\bar \theta}^2}  \ .
\ena
The correlation is defined in (\ref{eq:4DLangevin-eq1}). It reads 
\bea
\langle 
\Delta w^a | ( \t, x ) 
\Delta w^b |_{\theta^2{\bar \theta}^2} ( \t, x' ) 
\rangle_{\Delta w_\t} 
& = & 
\langle 
\Delta w^a |_{\theta^2{\bar \theta}^2} ( \t, x ) 
\Delta w^b | ( \t, x' )  
\rangle_{\Delta w_\t}       \nn 
& = &
2\Delta \t \delta^{ab}\delta ( x-x' )         \ , \nn 
\langle 
\Delta w^a |_{\theta^2}( \t, x ) \Delta w^b |_{{\bar \theta}^2} ( \t, x' ) 
\rangle_{\Delta w_\t} 
& = & 
\langle 
\Delta w^a |_{{\bar \theta}^2} ( \t, x ) \Delta w^b |_{\theta^2}( \t, x' ) 
\rangle_{\Delta w_\t}   \nn
& = & 
2\Delta \t \delta^{ab}\delta ( x-x' )      \ , \nn 
\langle 
{\Delta w}^a |_{\theta\ \alpha} ( \t, x ){\Delta w}^b |_{{\bar \theta}^2 \theta\ \beta } ( \t, x' ) 
\rangle_{\Delta w_\t}       
& = & 
\langle 
{\Delta w}^a |_{{\bar \theta}^2 \theta\ \alpha} ( \t, x ) {\Delta w}^b |_{\theta\ \beta} ( \t, x' ) 
\rangle_{\Delta w_\t}        \nn
& = & 
4\Delta \t \epsilon_{\alpha\beta} \delta^{ab}\delta ( x-x' )      \ , \nn
\langle 
\Delta w^a |_{{\bar \theta}}^{\dot \alpha} ( \t, x ) {\Delta w}^b |_{\theta^2 {\bar \theta}}^{\dot \beta}  ( \t, x' ) 
\rangle_{\Delta w_\t} 
& = & 
\langle 
{\Delta w}^a |_{\theta^2 {\bar \theta}}^{\dot \alpha}  ( \t, x ) \Delta w^b |_{\bar \theta}^{\dot \beta} ( \t, x' ) 
\rangle_{\Delta w_\t}    \nn 
& = & 
4\Delta \t \epsilon^{{\dot \alpha}{\dot \beta}} \delta^{ab}\delta ( x-x' )               \ , \nn 
\langle 
\Delta w^a |_{\theta\sigma^m{\bar \theta}} ( \t, x ) \Delta w^b |_{\theta\sigma^n{\bar \theta}} ( \t, x' ) 
\rangle_{\Delta w_\t} 
& = &  
- 4\Delta \t \eta^{mn} \delta^{ab}\delta ( x-x' )   \ . 
\ena
The hermitian conjugates of the noise component fields are determined from the condition (\ref{eq:Wess-Zumino-nise-B-eq2}).
In the almost Wess-Zumino gauge, the hermitian conjugate is trivial for 
$
\Delta w^\dagger | 
$, 
$
\Delta w^\dagger |_\theta 
$, 
$
\Delta w^\dagger |_{\bar \theta} 
$, 
$
\Delta w^\dagger |_{\theta^2}
$ 
and 
$
\Delta w^\dagger |_{{\bar \theta}^2} 
$.
The non-trivial components are given by
\bea
\Delta w^\dagger |_{\theta\sigma^m{\bar \theta}}
& = & 
\Delta w |_{\theta\sigma^m{\bar \theta}} 
+ 2 [ v_m,\ \Delta w | ]     \  , \nn
\Delta w^\dagger |_{\theta^2{\bar \theta}} 
& = & 
\Delta w |_{\theta^2{\bar \theta}} 
- 2 i [{\bar \lambda},\ \Delta w | ]  
+{\bar \sigma}^m [ v_m,\ \Delta w |_\theta ]  \  , \nn 
\Delta w^\dagger |_{{\bar \theta}^2\theta} 
& = & 
\Delta w |_{{\bar \theta}^2\theta} 
+ 2 i [\lambda,\ \Delta w | ]  
- \sigma^m [ v_m,\ \Delta w |_{\bar \theta} ]    \  , \nn 
\Delta w^\dagger |_{\theta^2{\bar \theta}^2} 
& = & 
\Delta w |_{\theta^2{\bar \theta}^2} 
- [ D,\ \Delta w | ] 
+ \eta^{mn}[ v_m,\ \Delta w |_{\theta\sigma^n{\bar \theta}} ]  \nn
& {} & \qquad 
+ i \{ {\bar \lambda},\  \Delta w |_{\bar \theta} \}   
- i \{ \lambda,\  \Delta w |_\theta \}     
- \eta^{mn}[ v_m,\ [ v_n,\ \Delta w | ] ]       \  . 
\ena
As is clear from the Langevin equations for the auxiliary component fields (\ref{eq:4DLangevin-c-eq2}), it is impossible to fix the gauge to the Wess-Zumino gauge, ${\tilde C} = {\tilde \chi} = {\bar {\tilde \chi}} = {\tilde M} = {\tilde N} = 0$. We note that there remains time development of the auxiliary component fields that is completely determined by the components in the Wess-Zumino gauge ($v_m$, $\lambda$, ${\bar \lambda}$, $D$). This is the reason that we call the superfield Langevin equation (\ref{eq:4DLangevin-B-eq5}) the \lq\lq  almost Wess-Zumino gauge \rq\rq. 

To obtain the Langevin equations for ($v_m$, $\lambda$, ${\bar \lambda}$, $D$) which include the auxiliary component fields, it is convenient to express (\ref{eq:4DLangevin-B-eq5}) as 
\bea
\label{eq:4DLangevin-c-eq3}
\Delta V|_{\rm WZ}                  
& = & 
 \displaystyle{\frac{L_V|_{\rm WZ}}{( 1 - e^{- 2L_V|_{\rm WZ}} )}} 
\Big( 
\Delta \t \displaystyle{\frac{1}{4g^2}} {\cal D}{\cal W}|_{\rm WZ}  
- \Delta \Theta  + \displaystyle{\frac{1}{2}} \Delta w^\dagger                  \Big)                           \nn 
& {} & \quad + 
\displaystyle{\frac{L_V|_{\rm WZ}}{(e^{2L_V|_{\rm WZ}} - 1 )}} 
\Big( 
\Delta \t \displaystyle{\frac{1}{4g^2}} 
{\bar {\cal D}}{\bar {\cal W}}|_{\rm WZ} 
- \Delta \Theta^\dagger + \displaystyle{\frac{1}{2}} \Delta w 
\Big)                                   \ .  
\ena
We note the covariant structure of the auxiliary component contributions in (\ref{eq:4DLangevin-c-eq3}) under the residual local gauge transformation ensures the residual local gauge covariance of the component Langevin equations. As a result of the definiton of the \lq\lq almost Wess-Zumino gauge\rq\rq, we can expand the expression for a superfield $F$ as 
\bea
 \displaystyle{\frac{L_V|_{\rm WZ}}{( 1 - e^{- 2L_V|_{\rm WZ}} )}} F     
& = & 
\displaystyle{\frac{1}{2}} F 
+ \displaystyle{\frac{1}{2}} [ V|_{\rm WZ},\ F ]                 
+ \displaystyle{\frac{1}{6}} [ V|_{\rm WZ},\ [ V|_{\rm WZ},\ F ] ] , \nn 
\displaystyle{\frac{L_V|_{\rm WZ}}{(e^{2L_V|_{\rm WZ}} - 1 )}} F         
& = & 
\displaystyle{\frac{1}{2}} F 
- \displaystyle{\frac{1}{2}} [ V|_{\rm WZ},\ F ] 
+ \displaystyle{\frac{1}{6}} [ V|_{\rm WZ},\ [ V|_{\rm WZ},\ F ] ]     \ . 
\ena

Substituting the component expansion in the Wess-Zumino gauge into the superfield Langevin equation (\ref{eq:4DLangevin-c-eq3}), we obtain the component expressions for $(v_m, \lambda, {\bar \lambda}, D)$: 
\bea
\Delta v_m
& = & 
- \displaystyle{\frac{\Delta\t}{2}} \displaystyle{\frac{1}{g^2}} ( 
{\cal D}^n v_{mn} + \lambda \sigma_m {\bar \lambda} + 
{\bar \lambda}{\bar \sigma}_m \lambda )  +  \Delta W_{v_m}    \ , \nn 
\Delta W_{v_m} 
& \equiv & 
- \displaystyle{\frac{1}{4}}\Delta w |_{\theta \sigma^m {\bar \theta}}    
- \displaystyle{\frac{1}{4}}\Delta w^\dagger |_{\theta \sigma^m {\bar \theta}}       \ , \nn     
\Delta \lambda 
+ \displaystyle{\frac{i}{2}} \sigma^m {\cal D}_m \Delta {\bar {\tilde \chi}} 
& = & 
- \displaystyle{\frac{\Delta\t}{2}} \displaystyle{\frac{1}{2 g^2}} 
\sigma^m {\bar \sigma}^n {\cal D}_m {\cal D}_n \lambda 
+ \Delta W_{\lambda}          \ , \nn 
\Delta W_{\lambda}  
& \equiv & 
\displaystyle{\frac{i}{4}} \Delta w |_{\theta {\bar \theta}^2} 
+ \displaystyle{\frac{i}{4}} \Delta w^\dagger |_{\theta {\bar \theta}^2}                      \ ,\nn 
\Delta {\bar \lambda} 
+ \displaystyle{\frac{i}{2}} {\bar \sigma}^m {\cal D}_m \Delta {\tilde \chi} 
& = & 
- \displaystyle{\frac{\Delta\t}{2}} \displaystyle{\frac{1}{2 g^2}} 
{\bar \sigma}^m \sigma^n {\cal D}_m {\cal D}_n {\bar \lambda} 
+ \Delta W_{\bar \lambda}          \ , \nn 
\Delta W_{\bar \lambda}  
& \equiv & 
- \displaystyle{\frac{i}{4}} \Delta w |_{\theta^2 {\bar \theta}} 
- \displaystyle{\frac{i}{4}} \Delta w^\dagger |_{\theta^2 {\bar \theta}}                \ ,\nn 
\Delta D + \displaystyle{\frac{1}{2}}{\cal D}^m {\cal D}_m \Delta {\tilde C} 
& = & 
\displaystyle{\frac{\Delta\t}{2}}  \displaystyle{\frac{1}{2 g^2}}{\cal D}^m {\cal D}_m D + \Delta W_D      \ , \nn
\Delta W_D 
& \equiv & 
\displaystyle{\frac{1}{2}} \Delta w |_{\theta^2 {\bar \theta}^2} 
+ \displaystyle{\frac{1}{2}} \Delta w^\dagger |_{\theta^2 {\bar \theta}^2}            \ .
\ena
Here we have used the Langevin equations for the auxiliary component fields. For example, in the Langevin equation for $D$, there appears the non-covariant expression  
$
\displaystyle{\frac{1}{6}} [ v^m,\ [v_m,\  
- \Delta {\tilde C} - \displaystyle{\frac{\Delta \t}{2}}\displaystyle{\frac{1}{g^2}} D + \displaystyle{\frac{1}{2}} \Delta w | 
] ], 
$
which vanishes with the help of the Langevin equation for $\Delta {\tilde C}$ in (\ref{eq:4DLangevin-c-eq2}). 

In order to obtain the Langevin equations for the components in the Wess-Zumino gauge, we use the Langevin equations for the auxiliary component fields. 
Finally, we obtain the Langevin equations for the components in the Wess-Zumino gauge: 
\bea
\Delta v_m
& = & 
- \displaystyle{\frac{\Delta\t}{2}} \displaystyle{\frac{1}{g^2}} ( 
{\cal D}^n v_{mn} + \lambda \sigma_m {\bar \lambda} + 
{\bar \lambda}{\bar \sigma}_m \lambda )  +  \Delta W_{v_m}    \ , \nn 
\Delta \lambda 
& = & 
- \displaystyle{\frac{\Delta\t}{2}} \displaystyle{\frac{1}{g^2}} 
\sigma^m {\bar \sigma}^n {\cal D}_m {\cal D}_n \lambda 
+ \Delta W_{\lambda} 
+ \displaystyle{\frac{1}{4}} \sigma^m {\cal D}_m \Delta w|_{\bar \theta}                   \ , \nn 
\Delta {\bar \lambda} 
& = & 
- \displaystyle{\frac{\Delta\t}{2}} \displaystyle{\frac{1}{2 g^2}} 
{\bar \sigma}^m \sigma^n {\cal D}_m {\cal D}_n {\bar \lambda} 
+ \Delta W_{\bar \lambda} 
- \displaystyle{\frac{1}{4}} {\bar \sigma}^m {\cal D}_m \Delta w |_\theta              \ , \nn 
\Delta D 
& = & 
\displaystyle{\frac{\Delta\t}{2}}  \displaystyle{\frac{1}{ g^2}}{\cal D}^m {\cal D}_m D + 
\Delta W_D 
- \displaystyle{\frac{1}{4}} {\cal D}^m {\cal D}_m \Delta w |    \ .
\ena
These Langevin equations actually reproduce the action for SYM$_4$ in the Wess-Zumino gauge. This fact can be confirmed by determining the kernel structure of these Langevin equations. The correlations for the noise component fields are given by 
\bea
& {} & 
\langle  
( \Delta W_{v_m} )_x^a ( \Delta W_{v_n} )_y^b 
\rangle_{\Delta w_\t}                     \nn 
& {} & \qquad 
 =  
 - \displaystyle{\frac{1}{2}} 2\Delta \t \delta^{ab} \eta_{mn}  \delta^4 ( x-y )                 \ , \nn
& {} & 
\langle  
( \Delta W_{\lambda} 
+ \displaystyle{\frac{1}{4}} \sigma^m {\cal D}_m \Delta w|_{\bar \theta} )_{x \alpha}^a 
( \Delta W_{\bar \lambda} 
- \displaystyle{\frac{1}{4}} {\bar \sigma}^m {\cal D}_m \Delta w |_\theta )_y^{b {\dot \beta}}
\rangle_{\Delta w_\t}                     \nn 
& {} & \qquad 
 = - \displaystyle{\frac{i}{2}} 2\Delta\t ( \sigma^m \epsilon )_\alpha^{\ {\dot \beta}} 
 ({\cal D}_m)_x^{\ ab}  \delta^4 ( x-y )                   \ , \nn
& {} & 
\langle  
( \Delta W_D 
- \displaystyle{\frac{1}{4}} {\cal D}^m {\cal D}_m \Delta w | )_x 
( \Delta W_D 
- \displaystyle{\frac{1}{4}} {\cal D}^m {\cal D}_m \Delta w | )_y 
\rangle_{\Delta w_\t}                     \nn 
& {} & \qquad 
 =  
 - \displaystyle{\frac{1}{2}} 2\Delta \t \delta^{ab}  
  {\cal D}^m {\cal D}_m \delta^4 ( x-y )                   \ .
\ena
This result shows that the kernels for the Langevin equations for $v_m$, $( \lambda, {\bar \lambda} )$ and  $D$ are given by 
$- \displaystyle{\frac{1}{2}}$, 
$- \displaystyle{\frac{i}{2}} ( \sigma^m \epsilon )_\alpha^{\ {\dot \beta}} 
 ({\cal D}_m)_x^{\ ab}$ 
 and 
$- \displaystyle{\frac{1}{2}}   
  {\cal D}^m {\cal D}_m $, 
respectively.

\end{document}